%% file: tileAgeingPaper.tex
\documentclass[cernpreprint, PAPER, cernlogo=true, atlaslogo=false, atlasdraft=false, coverpage=false, texlive=2023, UKenglish, texmf, orcidlogo]{atlasdoc}
\usepackage{atlaspackage}
\usepackage{atlasbiblatex}

\usepackage{atlasphysics}
\graphicspath{{logos/}{figures/}}

\AtlasTitle{Study of the Radiation Hardness of the ATLAS Tile Calorimeter Optical Instrumentation with Run 2 data}
\AtlasAbstract{%
This paper presents a study of the radiation hardness of the hadronic Tile Calorimeter of the ATLAS experiment in the LHC Run~2.
Both the plastic scintillators constituting the detector active media and the wavelength-shifting optical fibres collecting the scintillation light
into the photodetector readout are elements susceptible to radiation damage.
The dedicated calibration and monitoring systems of the detector (caesium radioactive sources, laser and
minimum bias integrator) allow to assess the response of these optical components. Data collected with these systems between 2015 and 2018 are
analysed to measure the degradation of the optical instrumentation across Run~2.
Moreover, a simulation of the total ionising dose in the calorimeter is employed to study and model the degradation profile
as a function of the exposure conditions, both integrated dose and dose rate. The measurement of the relative light output loss in Run~2
is presented and extrapolations to future scenarios are drawn based on current data. The impact of radiation damage on the cell response uniformity is also analysed.
}

\AtlasJournal{\JINST}
\AtlasJournalRef{2025 JINST 20 P06006}
\AtlasDOI{doi.org/10.1088/1748-0221/20/06/P06006}

\hypersetup{pdftitle={ATLAS document},pdfauthor={The ATLAS Collaboration}}

\begin{document}

\maketitle
\input{atlas_authlist}

\clearpage
\tableofcontents

\section{Introduction}
\label{sec:intro}

\newcommand{\AtlasCoordFootnote}{
ATLAS uses a right-handed coordinate system with its origin at the nominal interaction point (IP) in the centre of the detector and the \(z\)-axis along the beam pipe.
The \(x\)-axis points from the IP to the centre of the LHC ring, and the \(y\)-axis points upwards. Cylindrical coordinates \((r,\phi)\) are used in the transverse plane,
\(\phi\) being the azimuthal angle around the \(z\)-axis. The pseudorapidity is defined in terms of the polar angle \(\theta\) as \(\eta = -\ln \tan(\theta/2)\).
Angular distance is measured in units of \(\Delta R \equiv \sqrt{(\Delta\eta)^{2} + (\Delta\phi)^{2}}\).}

\newcommand{\pileupFootnote}{
Pile-up consists of the overlay of signals from multiple proton-proton interactions within the triggered bunch crossing and from neighbouring bunch crossings.}

The Tile Calorimeter (TileCal) is the central hadronic calorimeter~\cite{ATLAS-TDR-03,TCAL-2010-01} of the ATLAS experiment~\cite{PERF-2007-01} at the LHC~\cite{Evans:2008zzb}. This system provides measurements of hadrons, jets, hadronic decays of $\tau$-leptons and missing energy, also participating in muon identification and sending analogue signals to the hardware-based Level-1 calorimeter trigger~\cite{TCAL-2017-01,TCAL-2021-01}. The TileCal detector spans over the pseudorapidity range of $|\eta|<1.7$~\footnote{\AtlasCoordFootnote} and consists of a plastic scintillator tile (active medium) and low-carbon steel (absorber medium) sampling technology. The scintillation light readout is ensured by wavelength-shifting (WLS) optical fibres and vacuum photomultiplier tubes (PMTs). In total, the TileCal comprises about 460,000 scintillator tiles and 550,000 WLS fibres.

The light output of optical components (scintillators and WLS fibres) changes in time due to natural ageing and radiation exposure. The radiation hardness of the individual components was tested during the phase of the detector design, and the maximum impact of natural ageing in the scintillator light output was estimated to be -1\% per year~\cite{Abdallah:2007cca}. Radiation damage on plastic scintillators is known to affect both the scintillation process, reducing the intrinsic absolute light yield, and the light attenuation, increasing the signal loss within the material~\cite{ZORN1993377,LI2005449}.
Its fundamental mechanisms are well studied and identified with the formation of colour centres both affecting the material transparency and competing with the scintillating molecules.

The optical components of the TileCal were designed for 10~years of data-taking at the LHC considering the luminosity of $10^{34}\,\mathrm{cm}^{-2}\mathrm{s}^{-1}$. However, due to the foreseen High Luminosity (HL)-LHC phase, this detector will operate almost a two times longer period at higher instantaneous luminosity and increased pile-up\footnote{\pileupFootnote}~\cite{ZurbanoFernandez:2020cco}. The challenges inherent to the HL-LHC operation and technical solutions are discussed in the technical design report for the TileCal upgrade~\cite{ATLAS-TDR-28}. Whereas the PMTs might be exchanged, the scintillators and WLS fibres located inside the calorimeter volume cannot be replaced before the start of the HL-LHC programme. Studies of the TileCal optics robustness are crucial for understanding the detector's performance during the extended running period. First studies revealed evidence of a per cent-level degradation of scintillator light output in 2012 during the LHC Run~1, but only for the most exposed cell of TileCal~\cite{Fischer_2017}. This was later confirmed for an extended set of cells using the 2015--2017 partial dataset from Run~2~\cite{Pedro:2019lgd}. The full Run~2 data was then explored to measure and model the degradation of all TileCal cells, allowing preliminary extrapolations to future run conditions~\cite{BPereira}. Dose rate effects were also observed to have a significant impact on the scintillator degradation and were integrated into the model~\cite{CMSHCAL:2016dvd,TCAL-2021-01}.

This paper summarises the analysis of the TileCal optics robustness using Run~2 data, which corresponds to a delivered integrated luminosity of 157.4~\ifb~\cite{DAPR-2021-01}. The measurements of the changes in the response of the calorimeter in time were performed using dedicated calibration systems. The light yields are extracted and extrapolated to higher doses similar to the values expected for the HL-LHC run. The paper is organised as follows: the TileCal structure and the optical components are introduced in Section~\ref{sec:tilecal-optics}. The calibration systems used to assess the response of the TileCal optics are described in Section~\ref{sec:calibration}. The radiation environment faced by the calorimeter during the LHC Run~2 and the expected doses at the HL-LHC are discussed in Section~\ref{sec:radiation}. In Section~\ref{sec:ageing}, the study of the optics radiation damage and their expected robustness in future runs is presented and, finally, conclusions are drawn in Section~\ref{sec:conclusion}.


%

%
\section{Tile Calorimeter and Optical Instrumentation}
\label{sec:tilecal-optics}

\begin{figure}[t]
\centering
\subfloat[]{\includegraphics[width=0.585\linewidth]{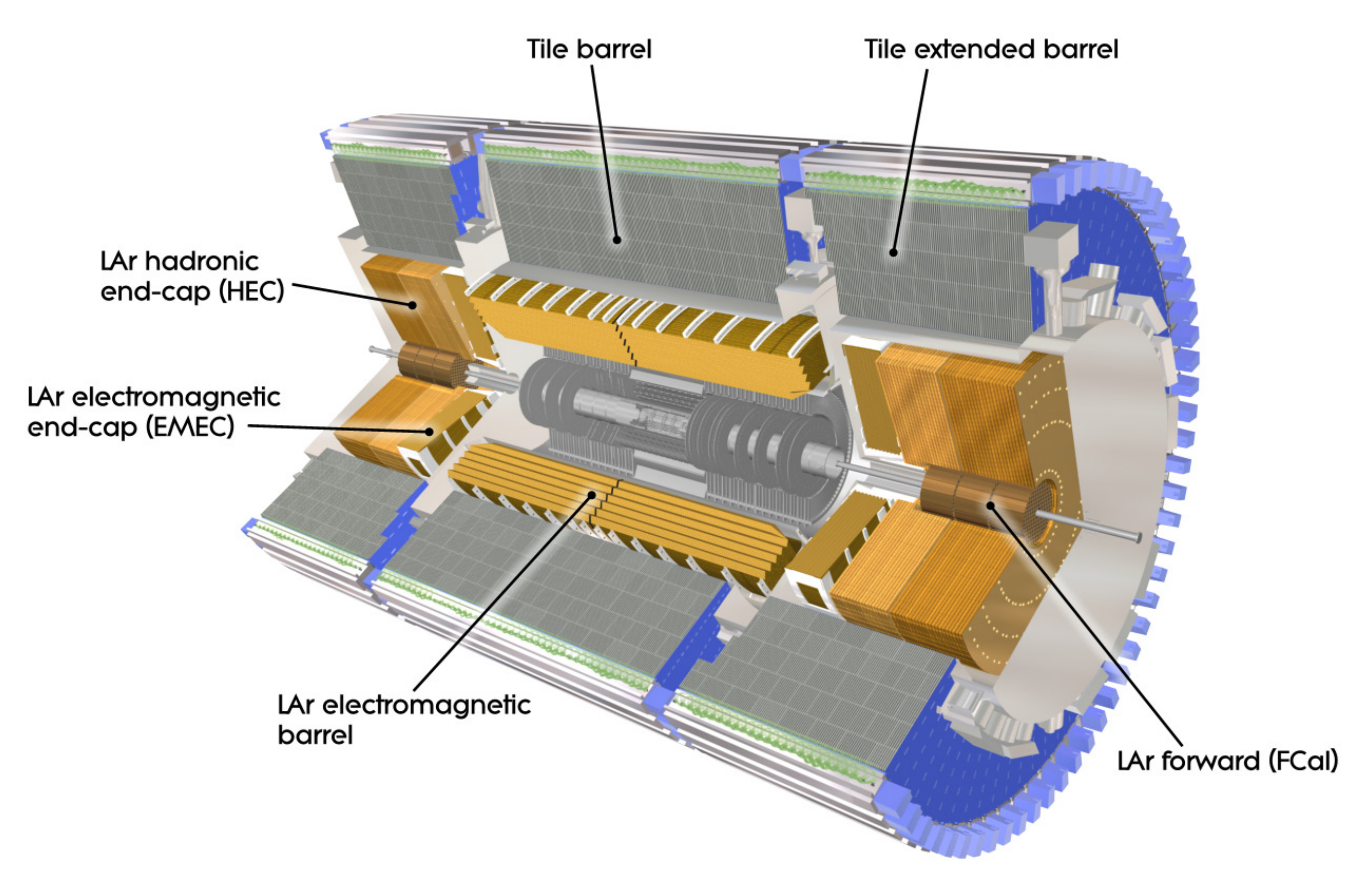}\label{fig:ATLAScalorimeters}}
\subfloat[]{\includegraphics[width=0.315\linewidth]{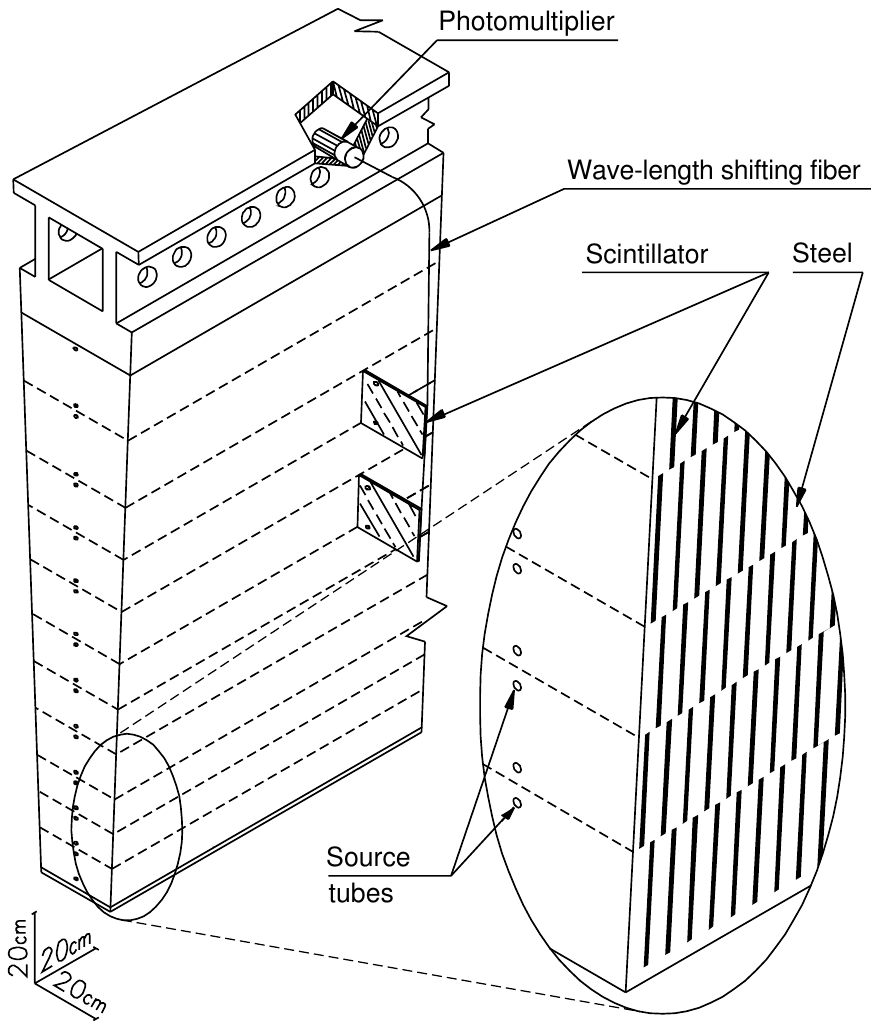}\label{fig:tile-readout}}
\caption{(a) Cut-away view of the ATLAS calorimeter systems. The TileCal consists of a barrel and two extended barrels. The sections of the ATLAS Liquid Argon (LAr) calorimeter are also indicated~\cite{TCAL-2010-01}. (b) Schema of the mechanical assembly and optical readout of a single TileCal module, exhibiting the radially staggered scintillators and steel plates, and the readout by WLS fibre and PMTs. In total, the modules comprise 11 rows in depth of scintillator/steel. Source tubes allow to circulate a $^{137}$Cs radioactive source for calibration purposes~\cite{TCAL-2010-01}.}
\end{figure}

The TileCal is a non-compensating sampling calorimeter, which design provides a standalone energy resolution for isolated pions of $\frac{\Delta E}{E}\sim \frac{56.4\%}{\sqrt E} \oplus 5.5\%$, measured at test beams~\cite{PERF-2007-01}.
It consists of a central Long Barrel (LB) and two Extended Barrels (EB) in the end-cap region, arranged as cylindrical layers surrounding the ATLAS Liquid Argon (LAr) calorimeter system as shown in Figure~\ref{fig:ATLAScalorimeters}~\cite{Abdallah:2013pea}. The barrels are divided into four logical partitions: LBA and EBA for $\eta>0$, and LBC and EBC for $\eta<0$. Each is composed of 64 independent wedge-shaped modules, defining an azimuthal granularity of $\Delta\phi\approx 0.1$. In each module, scintillating plastic tiles and low-carbon steel plates are interleaved, in a 1:4.7 density proportion, and radially staggered transversely to the LHC beam axis into 11 depth rows. The scintillator tiles are coupled to WLS fibres in two edges, which transport the light to the PMT readout. This arrangement is displayed in Figure~\ref{fig:tile-readout}.

The calorimeter cell units are defined by the common readout of groups of several WLS fibres, matching a set of scintillator tiles in a specific detector region. Since the tiles have WLS fibres coupled to both edges, each cell has double readout by two independent channels providing redundancy on the energy measurement. The LB (EB) modules are segmented into three radial cell layers - A, BC and D (A, B/C and D) - to sample hadronic showers in the longitudinal direction, and have lateral granularity of $\Delta\eta=0.1$ and $\Delta\eta=0.2$ in the two innermost layers (A, BC/B) and the outermost layer (D), respectively. Figure~\ref{fig:tilecal_cell_map} shows the TileCal cell layout for $\eta > 0$. In total, the TileCal comprises 5182 cells and 9852 PMT readout channels.

The D4, C10 and E1--E4 cells are the intermediate TileCal cells and provide coverage in the $0.8<|\eta|<1.6$ region between the LB and the EB, and information on energy losses in non-active material from other ATLAS sub-systems. A few D4 and C10 cells have reduced thickness or special geometry to accommodate services and readout electronics from other systems~\cite{ATLAS-TDR-03,Abdallah:2013pea}, while the gap (E1/E2) and crack (E3/E4) cells are composed of a single large scintillator and are exceptionally read by a single PMT.

\begin{figure}[t]
\centering
\includegraphics[width=0.8\textwidth]{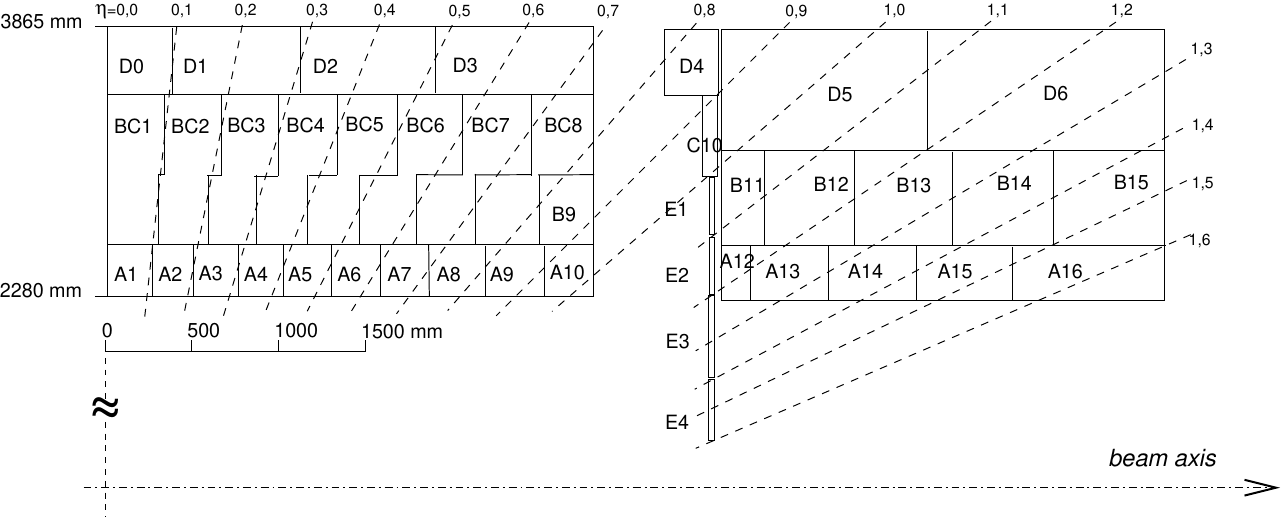}
\caption{TileCal cell layout for $\eta > 0$. The A-layer is the closest to the beamline. The naming convention is repeated for cells with $\eta<0$. The Long Barrel (Extended Barrel) cells are shown at the left (right)~\cite{ref:tilemap}. %
}
\label{fig:tilecal_cell_map}
\end{figure}

The scintillator plates composing the calorimeter barrels were manufactured by the SIA Luch Russian company in collaboration with the IHEP-Protvino TileCal group~\cite{Abdallah:2007cca,Abdallah:2009zza}. They are 3~mm thick and have a trapezoidal shape with different dimensions depending on the tile row: the width varies from 219.1~mm (row 1 at innermost radial position) to 368.2~mm (row 11 at outermost radial position), and the height also increases with the tile row radius as indicated in Table~\ref{tab:tileRowsHeights}.
The injection moulding technique was employed to obtain the different tile geometries from a mixture of optically transparent Polystyrene (PS) granules with a low mass concentration of fluorescent dyes (1.5\% of p-Terphenyl and 0.04\% of POPOP). The base PS granules were obtained from two manufacturers and are dubbed PSM115 and BASF165H. Approximately half the scintillators were produced from PSM115 and the other half from BASF165H, but 95\% of the modules were instrumented with PSM115 tiles in the tile rows 1, 2 and 3 (A-layer). Since the light yield differed between the two PS types, with the BASF165H having a larger average yield, this choice allowed to maximize the response uniformity in the innermost layer, where the energy deposits are larger.

\begin{table}[t]
\caption{Height of the TileCal scintillator tiles as a function of the tile row~\cite{Abdallah:2007cca}.}
\label{tab:tileRowsHeights}
\begin{center}
\begin{tabular}{|l|c|}
\hline
Tile rows & Height [mm]\\ \hline
1, 2, 3 & 97 \\
4, 5, 6 & 127 \\
7, 8, 9 & 147 \\
10, 11 & 187\\ \hline
\end{tabular}
\end{center}
\end{table}

The produced scintillators were quality controlled for light response and transmission properties, and tested for radiation resistance.
Figure~\ref{fig:tile_LRestimate} shows the distribution of a light yield quality estimator $(I_{x0}\cdot I_{x1})^{1/2}$, where $I_{x0}$ ($I_{x1}$) is the light signal produced by a $^{90}$Sr positioned in one tile edge (opposite tile edge), read by an optical fibre. This estimator is a proxy for the tile light response uniformity, by combining information from light yield ($I_{x0}$) and transmission ($I_{x1}/I_{x0}$)~\cite{Abdallah:2009zza}. Tile packs were sorted in $(I_{x0}\cdot I_{x1})^{1/2}$ and distributed in the detector to minimise the spread of $(I_{x0}\cdot I_{x1})^{1/2}$ within partitions, as shown in Figure~\ref{fig:tile_LRestimate}, maximising cell response uniformity.
Irradiation tests to bare scintillators were conducted in laboratory to characterise radiation resistance. Figure~\ref{fig:tile_irradiation_lab} shows the relative light response $I/I_0$ measured one month after irradiation with hadrons and gammas from a $^{137}$Cs radioactive source at $2-3\times 10^{-2}$~Gy/s and $6\times 10^{-2}$~Gy/s, respectively, for different total ionising doses~\cite{Karyukhin:1996aya,Abdallah:2007cca}. The dose after the nominal 10-year period of LHC operation was expected to be 360~Gy for the tiles facing the most severe energy density fluxes. According to the study, a maximum degradation of about 10\% was reached, assuming the recovery profile achieved with regular irradiation breaks of one month.

\begin{figure}[t]
\centering
\subfloat[]{\includegraphics[height=.4\textwidth]{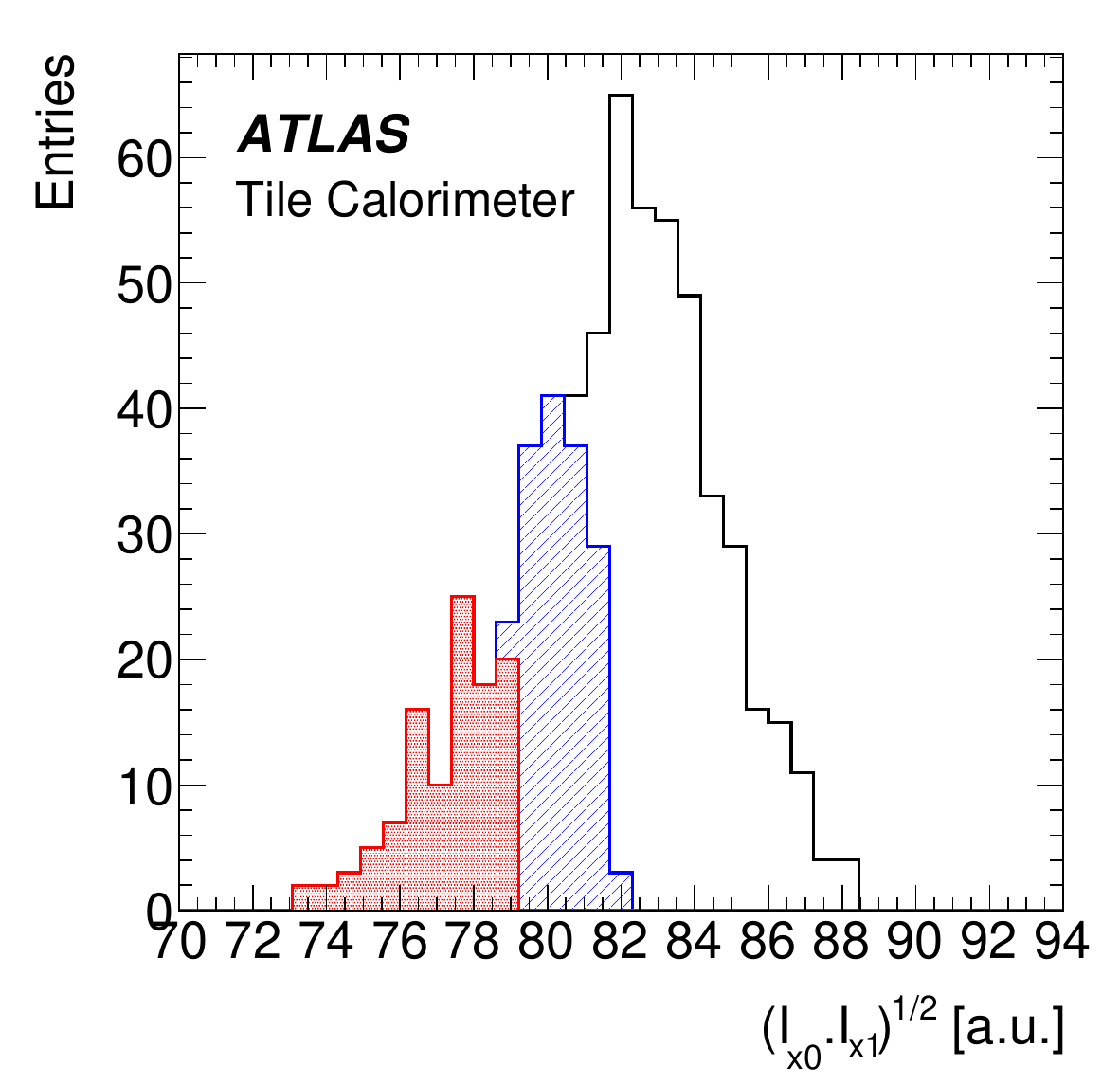}\label{fig:tile_LRestimate}}
\subfloat[]{\includegraphics[height=.38\textwidth]{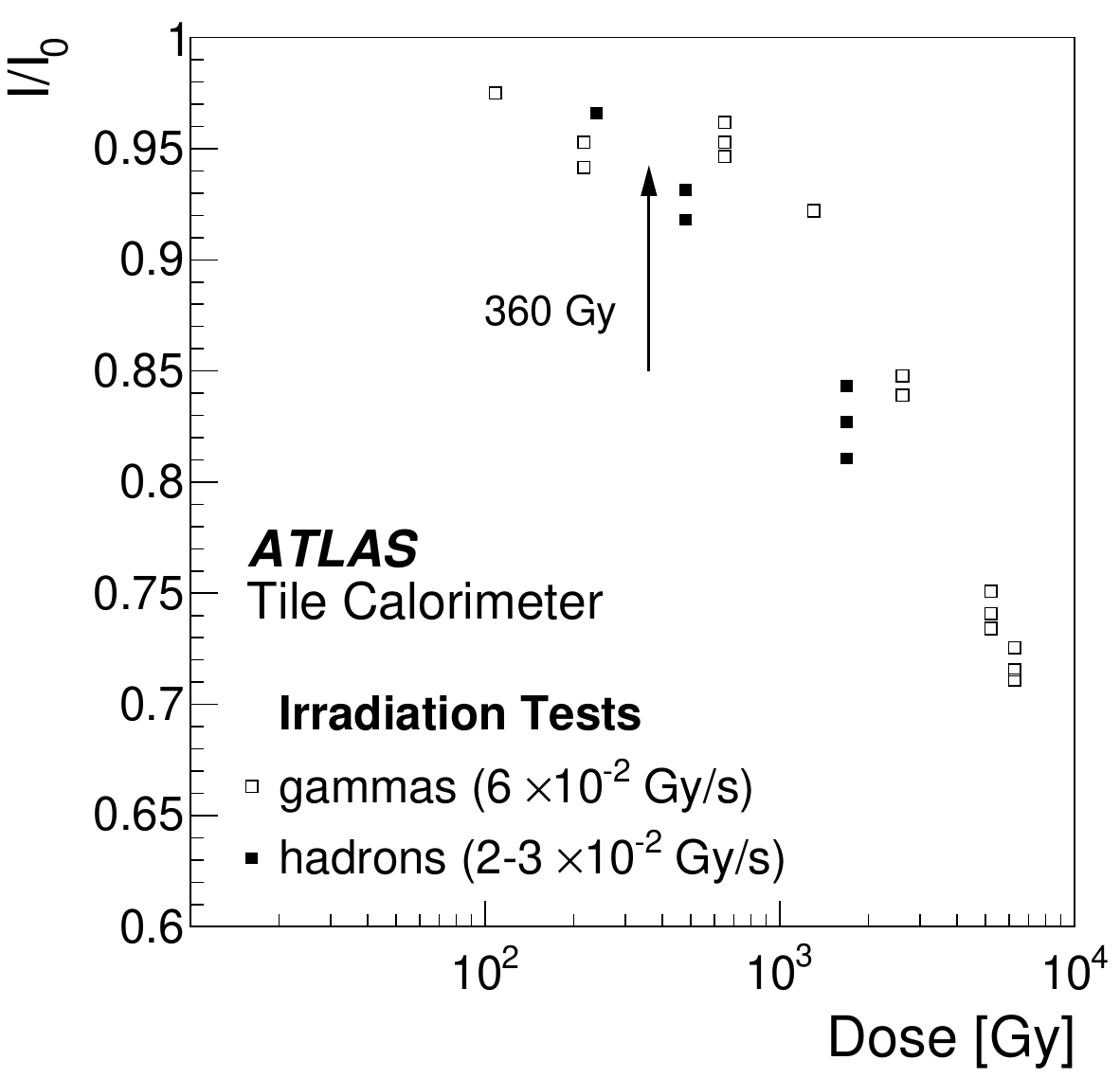}\label{fig:tile_irradiation_lab}}
\caption{(a) Distribution of $(I_{x0}\cdot I_{x1})^{1/2}$ for the tile packs used in EBA (red), EBC (blue) and LB (white and grey)~\cite{Abdallah:2009zza}.
(b) Relative light signal $I/I_0$ as a function of the dose measured one month after irradiations with hadrons and gamma rays, where the gamma rays are emitted from a $^{137}$Cs radioactive source, at dose rate $2-3\times 10^{-2}$~Gy/s and $6\times 10^{-2}$~Gy/s, respectively~\cite{Karyukhin:1996aya,Abdallah:2007cca}. The vertical arrow represents the expected dose of 360~Gy after the nominal 10-year period of LHC operation and points to the region where the respective $I/I_0$ is expected taking into account the data trend.}
\label{fig:tile_optics}
\end{figure}

In the TileCal construction, all scintillators were wrapped in Tyvek{\sffamily\textregistered} sleeves to improve light collection and protect the tile surfaces during assembly~\cite{Abdallah:2009zza}. The WLS fibres instrumenting the TileCal are 1~mm diameter double-clad Y11(200)MSJ produced by Kuraray~\cite{Burdin:1999cja,David:2002kva}. They are aluminised in the end opposite to the PMT to improve light collection~\cite{1312047}. Fibres reading different cells have different lengths, ranging from 88 (65)~cm for the shortest fibre collecting light from the D-layer, to 214 (232)~cm for the longest one coupling to the A-layer in the LB (EB). The overall module response uniformity, defined as the average RMS/mean signal induced by a $^{137}$Cs source in the cell tiles, was measured to be 5 to 8\% for the LB modules and up to 9\% for the EB modules, which is within the 10\% specification in both cases~\cite{Abdallah:2009zza}. Details of the study can be found in Ref.~\cite{Abdallah:2009zza}. Test beams demonstrated that the average light yield of the TileCal module arrangement is 72~p.e./GeV and 91~p.e./GeV for cells made with PSM115 and BASF165H scintillators, respectively, implying a sufficient signal-to-noise ratio for muon detection.

The scintillators of the E1 and E2 (E3 and E4) cells are 12~(6)~mm thick and made from a different material than the standard TileCal barrel ones: the commercial BC-408 (polyvinyl toluene-based) from Bicron was employed in E1 and E2, and the Ukrainian UPS-923A~\cite{Artikov:2005mg} (PS-based doped with 2\% PTP and 0.03\% POPOP) in E3 and E4~\cite{Pedro:2019lgd}. %

The Minimum Bias Trigger Scintillators (MBTS) system~\cite{Sidoti_2014}, mainly used for triggering purposes, is also read out by the TileCal EB electronics. It consists of 2~cm thick scintillator counters made of the same material as the E3/E4 crack cells - UPS-923A (PS with 2\% PTP and 0.03\% POPOP). The system comprises eight inner counters ($2.76 \leq |\eta| < 3.86$) and eight outer counters ($2.08 \leq |\eta| < 2.76$).%


%

%
\FloatBarrier
\section{Readout electronics, Calibration and Calorimeter response in Run 2}
\label{sec:calibration}

The PMTs and front-end electronics of each module are housed in steel girders and put in aluminium units that can be fully extracted. The nominal number of active channels per module is 45 and 32 in the LB and EB, respectively.

Each channel consists of a unit called a PMT block, which contains the light-mixer, the PMT tube and high-voltage (HV) divider, and a so-called 3-in-1 card~\cite{Anderson:2005ym}. This card is responsible for fast signal shaping, and the slow integration of the PMT signal. It also provides an analogue trigger to a sum board allowing to determine the total energy in a calorimeter tower within $\Delta \eta \times \Delta \phi = 0.1 \times 0.1$ to be used in the ATLAS trigger system~\cite{ATLASTDAQ:2016pov}. The fast PMT signal is shaped with 50~ns shaping time and sent through two linear outputs with a nominal relative gain of 64. These low and high gain signals are digitized, sampled and stored in pipelines at the LHC bunch crossing rate of 40~MHz using two 10-bit Analogue-to-Digital Converters (ADCs), hence forming an overall 16-bit dynamic range. Upon reception of a Level-1 trigger \cite{TRIG-2019-04}, seven readout samples centred around the bunch crossing ID of the trigger signal are packed and sent to the off-detector readout
electronics. An integrator readout with 6 switchable gains and a typical integration time of 10\,ms receives a small fraction of the PMT current and produces a digitized output through a 12-bit ADC~\cite{GonzalezParra:2011cla}. It is used for the calibration of the calorimeter response using caesium signal during dedicated scans, and also for measuring the current generated by minimum bias (MB) proton--proton ($pp$) collisions~\cite{DAPR-2013-01}.

A dedicated calibration system is used to monitor the detector components at each step of the detector readout.
A movable caesium radioactive $\gamma$-source scans the optical components and the PMTs in each barrel cylinder. The laser system monitors the PMTs and the digital readout electronics, and the charge injection system calibrates the front-end electronics. Figure~\ref{Fig:TileCalCalibrationChain} shows a flow diagram summarising the different calibration systems along with the paths followed by the signals from different sources. The systems are used to calibrate the different aspects of the detector signal response, defined as the reconstructed shaped and integrated signals to a given signal source. Furthermore, the information provided by the laser system, and the integrated signals from Cs scans and MB events, is complementarily explored to monitor the optics degradation of the different cells and MBTS counters.

\begin{figure}[t]
\centering
\includegraphics[width=1.\textwidth]{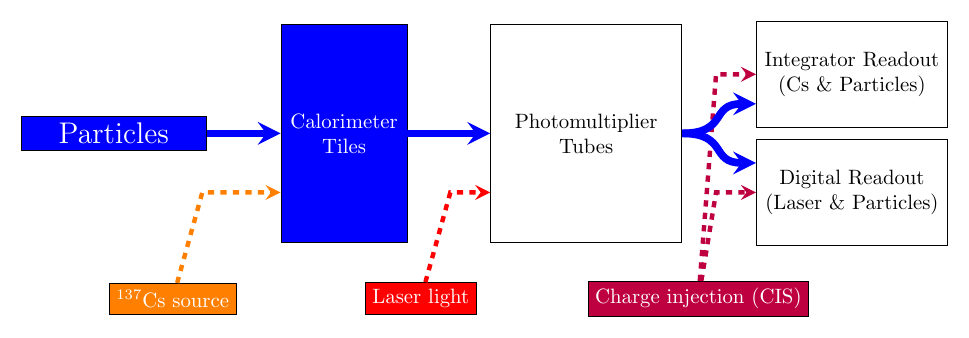}
\caption{The paths for each signal read out by the TileCal. The physics signal is denoted by the thick solid line and the path taken by each of the dedicated calibration systems is shown with dashed lines~\cite{TCAL-2021-01}.}
\label{Fig:TileCalCalibrationChain}
\end{figure}

\subsection{Caesium calibration}
\label{subsec:caesium_calibration}

The TileCal exploits radioactive $^{137}$Cs sources to maintain the global electromagnetic scale and to monitor the overall response of each readout channel~\cite{Cesium_2020}. A hydraulic system moves three sources through the calorimeter using a network of stainless steel tubes running through small holes in each tile scintillator. The beta decay of the $^{137}$Cs source produces a metastable nuclear isomer of $^{137m}$Ba. The isomer decays into the ground state by an emission of a 0.662~MeV photon which generates scintillation light in each tile. The average source activity was around 323~MBq at the beginning of Run~2 and decreases at an approximate rate of 2.3\% per year. To collect a sufficient signal, the electrical readout of the caesium calibration is performed using the integrator readout path. The integrator response is a measure of the integrated current in a PMT (integration time equals to 13.9\,ms).

The deviation of the TileCal channels response to the $^{137}$Cs source, $\Delta R_{\mathrm{Cs}}$, is monitored over time with respect to the beginning of Run~2.
Since the output signal of TileCal cells to the $^{137}$Cs source probes both the responses of the PMTs and the scintillating tiles, this monitoring system is key to study the evolution of the optics degradation across the years of detector operation. However, given that scanning the detector partitions with the $^{137}$Cs sources takes several hours, caesium data are only taken in long periods without LHC collision, either in technical stops or extended breaks. In total, there were around 25 scans during Run~2.

To determine $\Delta R_{\mathrm{Cs}}$, the complementary \emph{integral} and \emph{amplitude} methods are used~\cite{Cesium_2020}. They are illustrated in Figure~\ref{fig:example_cs_row_resp} showing the typical response of the third row of an A-cell as a function of the position of the $^{137}$Cs source relative to the tiles in the cell. In the integral method, a mean response $\langle \mathrm{Int} \rangle$ of the tile row is obtained by adding all the responses measured in the three regions labelled S0, S1 and S2, and normalising it to the width of the distribution, denoted by T2-T1, the notation is shown in Figure~\ref{fig:example_cs_row_resp}. The precision of these determinations for a typical cell is 0.3\%. In the amplitude method, the signal amplitude $A$ from every single tile in a row is obtained from a global fit to the row response with a precision of about 2~\%. The fit parametrises the individual tile response as traversed by the source by a sum of a Gaussian and an exponential function. %
In this case, the mean row response of the tile row $\langle A \rangle$ is obtained by averaging over the amplitudes $A$ of the tiles in the row.

\begin{figure}[t]
\centering
\includegraphics[width=0.5\textwidth]{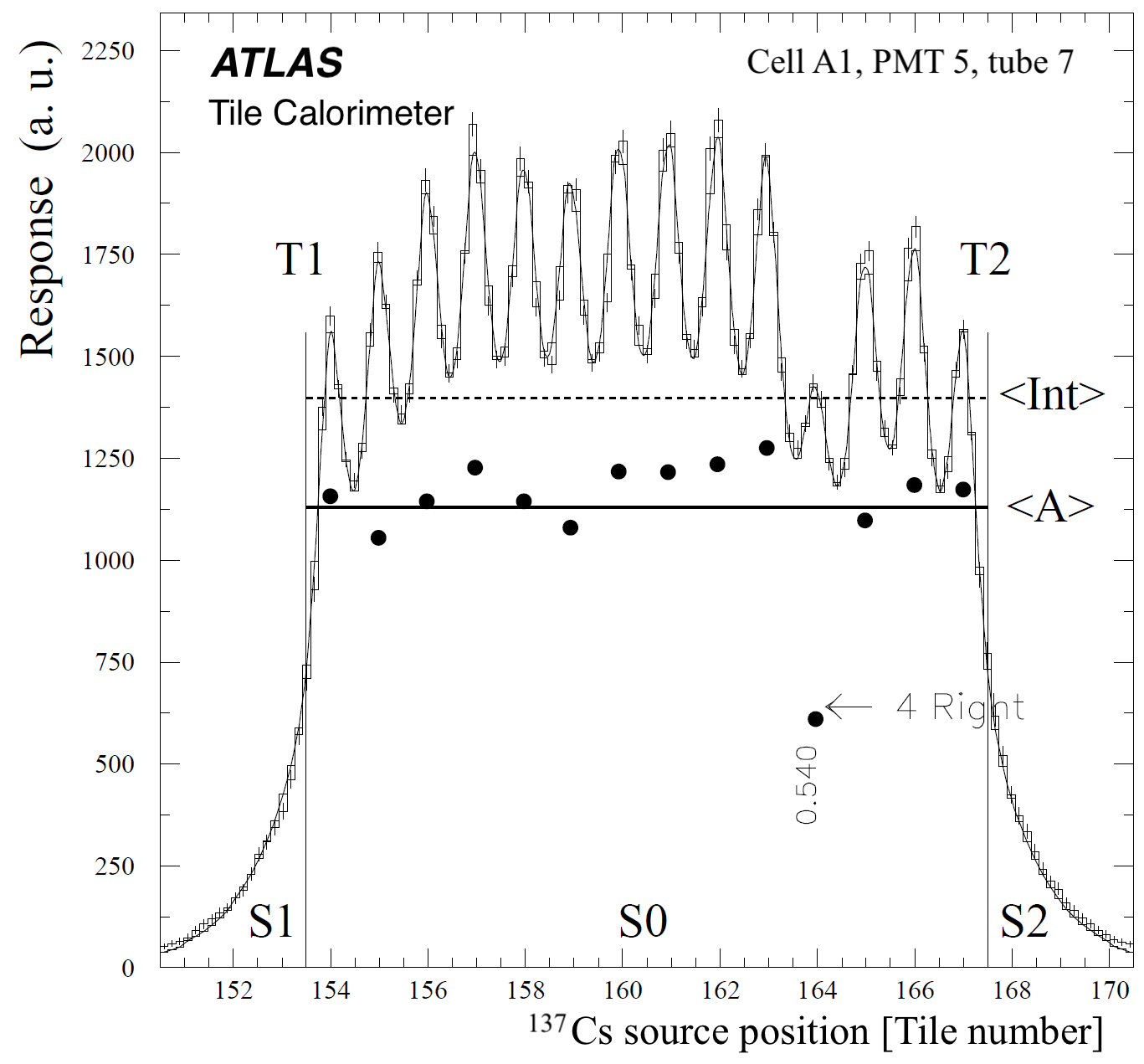}
\caption{Example of the sequencial response of a tile row from an A-cell taken at the 90~Hz readout frequency and presented as a function of the tile number where the
$^{137}$Cs source is positioned during the scan~\cite{Cesium_2020}.
The line labeled $\langle \mathrm{Int} \rangle$ indicates the mean response of the row calculated with the \emph{integral} method and the line labeled $\langle A \rangle$ is the mean response obtained with the \emph{amplitude} method~\cite{Cesium_2020}. Black points represent the individual tile amplitudes $A$ obtained through a global fit of the tile row response that parametrises each tile response with a sum of a Gaussian and an exponential function. As discussed in the text S0, S1 and S2 indicate the regions where the responses are measured. T1 and T2 correspond to the measured edges of the cell. PMT 5 stands for PMT number~5 and Tube 7 indicates the tile row number. The label ``4 Right'' indicates a tile with a bad fibre or a bad tile-fibre coupling.}
\label{fig:example_cs_row_resp}
\end{figure}

In both methods, the cell's response is found by averaging the mean response of its rows. Since the amplitude method is CPU-intensive, the integral method is employed for calibration and optics ageing studies at the cell level. On the other hand, the amplitude method is particularly useful for detailed optics ageing studies and for quantifying the uniformity of the cell response. These studies are presented in Section~\ref{sec:ageingCells}.

\subsection{Laser calibration}
\label{subsec:laser_calibration}

A laser calibration system is used to monitor the variations in the PMT response between caesium scans and the channel timing during collision data-taking periods~\cite{LaserRun2,Abdallah_2016}. The system consists of a single laser source located in a counting room approximately 100~m away from the detector, able to produce controlled short light pulses, which are simultaneously distributed by optical fibres to the photocathode in all 9852 PMTs. For Run~2, the laser system was improved for more long-term reliability~\cite{Gris:2217913}.

The deviation of the signal in each channel produced by laser events is measured concerning its nominal value at the start of the Run~2 data-taking, $\Delta R_{\mathrm{Las}}$. This allows to monitor the PMTs response with a total uncertainty of 0.5\% plus a luminosity-dependent sub-dominant term~\cite{LaserRun2}.

Laser calibration data are usually taken daily for both the gains in the absence of colliding beams, enabling frequent monitoring of the PMTs response evolution and important inputs for channel quality inspection. Laser pulses are also sent during empty bunch crossings of the LHC, providing data for monitoring the stability of the time calibration.

\subsection{Minimum bias integrator current calibration}
\label{subsec:minimum_bias}

Minimum bias inelastic $pp$ interactions at the LHC produce measurable signals in all the PMTs, which are used to monitor the variations of the calorimeter response over time using the integrator readout (as used by the caesium calibration system). The generated current is proportional to the instantaneous luminosity and produces signals in all cells of the calorimeter. It is used for calibration purposes and monitoring of the instantaneous luminosity of the colliding beams~\cite{DAPR-2021-01}.

The integrator PMT current ($I$) is measured as an increase of the ADC voltage relative to a pedestal value ($\mathrm{ped}$) measured just before the start of collisions, and normalised to the resistance of the integrator gain ($\mathrm{Int.\ gain}$) as follow:

\begin{equation}
I\ \mathrm{[nA]=\frac{ADC\ [mV] - ped\ [mV]}{Int.\ gain\ [M\Omega]}}
\label{eq:mb}
\end{equation}

The deviation of the integrator PMT current relative to the value at the beginning of Run~2 in each channel, $\Delta R_{\mathrm{MB}}$, is determined. To do so, the MB currents are first normalised to the track-counting measured by the inner detector, factoring out their direct dependence on the instantaneous luminosity~\cite{DAPR-2021-01}. The precision of the measurements is approximately 1.2\%. %
The measurements of $\Delta R_{\mathrm{MB}}$ monitor the changes in the response of the complete optical chain (scintillators, WLS fibres and PMTs) of TileCal.

\subsection{Calorimeter response variation}

The calorimeter response variation is monitored during Run~2 by the caesium and laser calibration systems as well as by the usage of the minimum bias events. Whereas the laser system is only sensitive to the changes in the PMT response, the caesium and minimum bias systems measure the variations caused by the changes in the response of the scintillating tiles, WLS fibres and PMTs. During $pp$ collisions, the response of most PMTs drifts down at different rates and the light output of scintillating tiles decreases due to radiation damage. During periods without collisions, the response for all cells recovers primarily due to an intrinsic restoration of the PMTs gain and potential annealing of scintillators.

\subsubsection{Standard calorimeter cells}

\begin{figure}[t]
\centering
\subfloat[]{\includegraphics[width=0.52\linewidth]{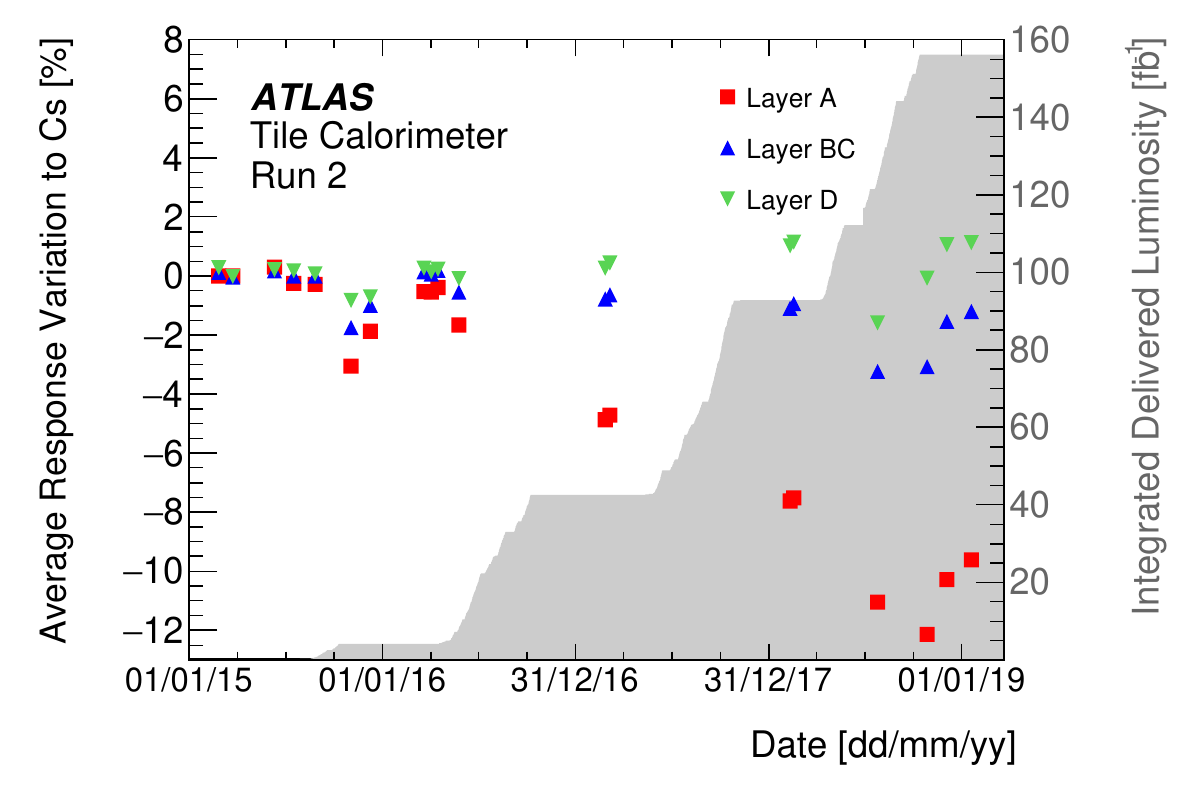}\label{fig:cesium_driftL}}
\subfloat[]{\includegraphics[width=0.475\linewidth]{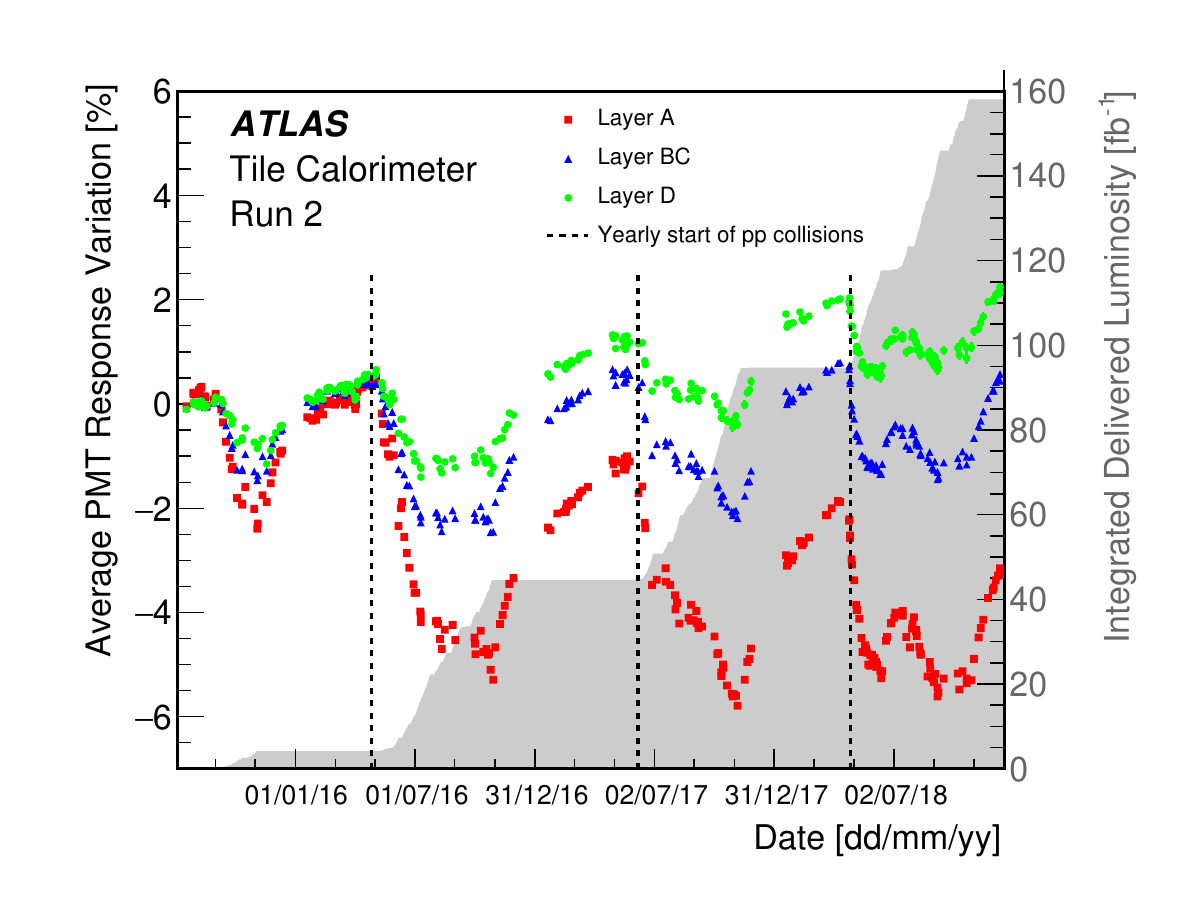}\label{fig:laser_evolution_run2}}
\caption{(a) The average of the response variation of the TileCal cells to the $^{137}$Cs source relative to the value measured at the beginning of Run~2, $\Delta R_{\mathrm{Cs}}$, as a function of time. The average runs over all cells in three radial layers. The increasing response corresponds to the periods without collisions~\cite{TCAL-2021-01}. (b) The mean response variation, $\Delta R_{\mathrm{Las}}$, in the PMTs for each layer, as a function of time, observed during the entire Run~2 (between standalone laser calibration data taken on 17 July 2015 and 22 October 2018)~\cite{TCAL-2021-01}. For each layer, the response variation is defined as the mean of a Gaussian function fit to the response variations in the channels associated with given layer. Known faulty channels are excluded. The LHC integrated delivered luminosity is shown by the shaded area~\cite{DAPR-2021-01}.}
\label{fig:cesium_laser_drift}
\end{figure}

Figure~\ref{fig:cesium_driftL} shows the deviations of the cell response to the $^{137}$Cs source from the value measured at the beginning of Run~2, $\Delta R_{\mathrm{Cs}}$, as a function of time, averaged over all cells in a given radial layer. Known faulty channels identified by data quality checks~\cite{TCAL-2021-01} are excluded from the cell average. Most affected cells are located at the inner radius of the calorimeter (A-layer). These cells obtain the highest irradiation doses which leads to the biggest drop in the response. Moreover, the received radiation dose, and therefore the degradation of the response, is not uniform across pseudorapidity. At the end of Run~2, the most irradiated cells in the A-layer had their response decreased by 18\%, while central cells in the outer D-layer drifted up by 2\%. The responses for all cells started to drift upward after the end of $pp$ collisions in October 2018 with a rate of about 1\% per month as seen in Figure~\ref{fig:cesium_driftL}.

Figure~\ref{fig:laser_evolution_run2} shows the time evolution of the mean gain variations in the PMTs, $\Delta R_{\mathrm{Las}}$,
for each layer observed during the entire Run~2. When a PMT is initially exposed to light after a long 'no light' period, its gain decreases rapidly and then slow stabilisation occurs. The recovery periods, in which the PMT response to the laser tends towards the initial conditions, coincide with the periods in which the LHC was not colliding protons. Data taking in each year started after periods of inactivity. The maximum average gain variation of --6\%  is observed in the PMTs from cells located in the A-layer at the end of $pp$ data-taking in Run~2. The response of the PMTs raised afterwards during the heavy ion data-taking and the technical stop to the mean variation of --3\% in the A-cells.

\subsubsection{Gap/Crack Scintillators}

\begin{figure}[t]
\centering
\subfloat[]{\includegraphics[height=0.38\linewidth]{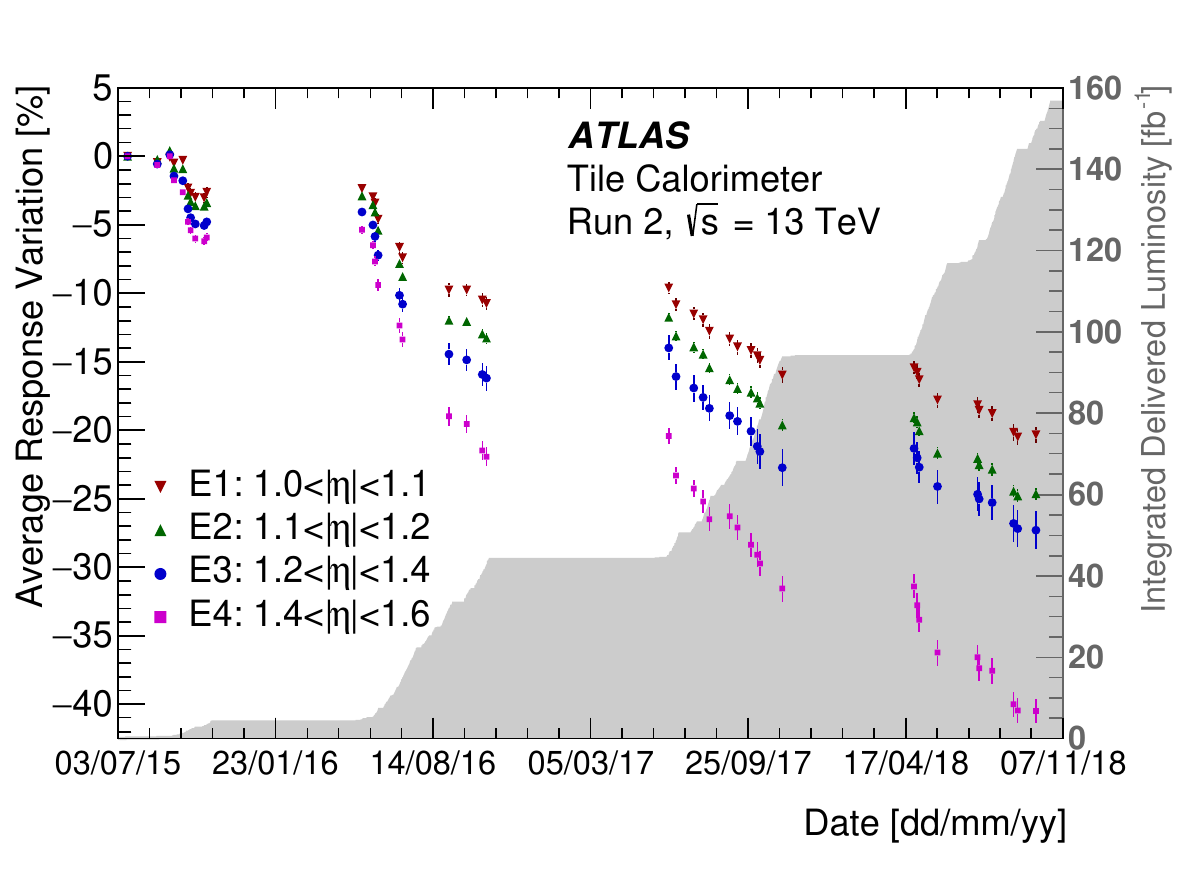}\label{fig:Ecell_mb}}
\hfill
\subfloat[]{\includegraphics[height=0.38\linewidth]{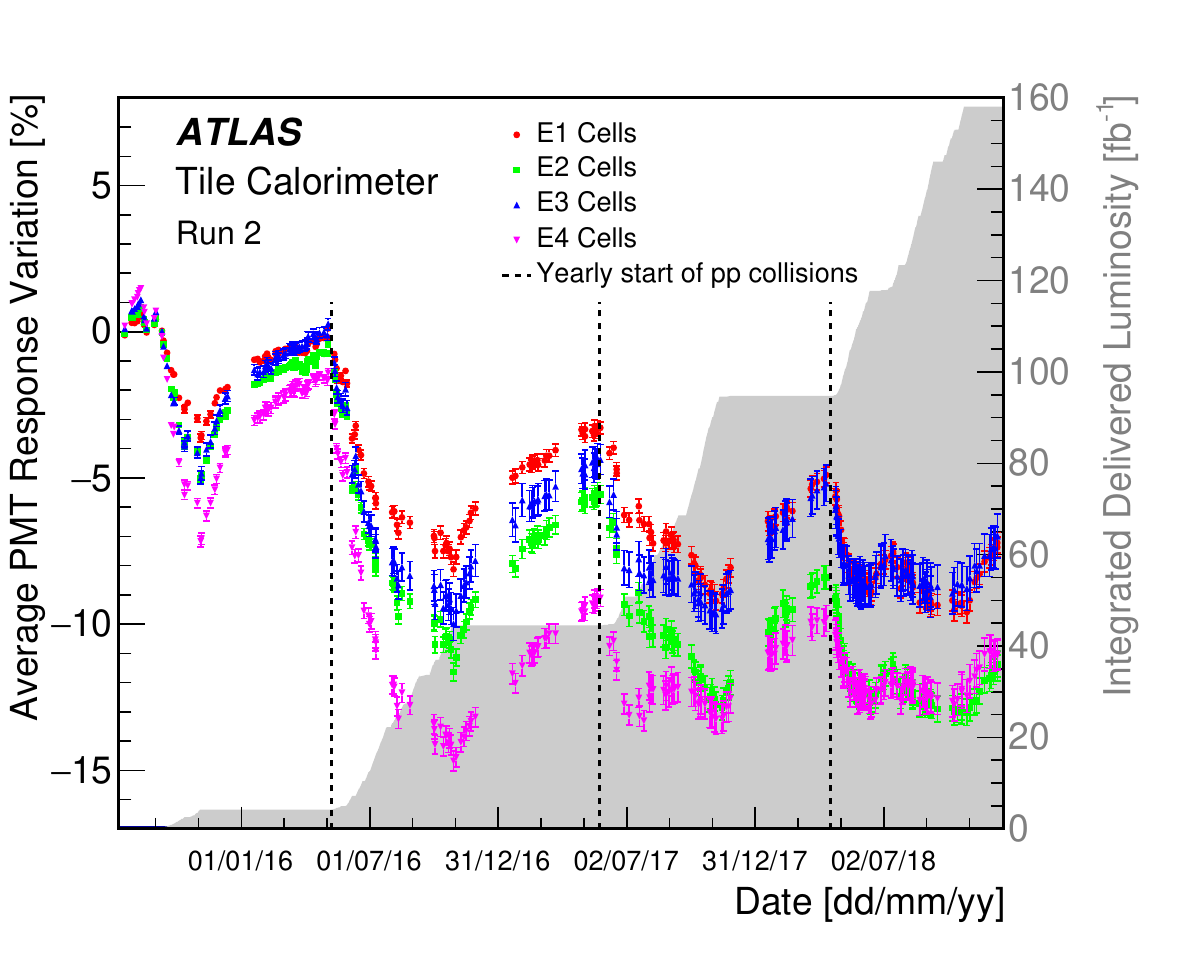}\label{fig:Ecell_laser}}
\caption{
(a) The variation of the average response to MB events, $\Delta R_{\mathrm{MB}}$, for the cells in the gap/crack region of the Extended Barrel as a function of time. The response to the MB events is normalised to the track-counting luminosity measured by the inner detector~\cite{DAPR-2021-01}. The vertical bars correspond to the RMS of all the response distributions. The results are normalised to the values measured in the first run of Run 2 (16th July 2015)~\cite{TCAL-2021-01}.
(b) The mean response variation, $\Delta R_{\mathrm{Las}}$, of the PMTs reading E1, E2, E3 and E4 cells, as a function of time, observed during the entire Run~2. For each cell, the response variation is defined as the mean of a Gaussian function fit to the response variations in the channels associated with given cell. Known faulty channels are excluded. The LHC integrated delivered luminosity is shown by the shaded area~\cite{DAPR-2021-01}.}
\label{fig:Ecell_drift}
\end{figure}

Minimum bias currents are used to measure the response of the cells not instrumented by caesium (e.g. cells in the gap and crack region of the EB, so-called E-cells, and MBTS counters) and calorimeter cells that show larger response variations during LHC operation. Figure~\ref{fig:Ecell_mb} shows the variation of the average response as seen by the minimum bias system for the E-cells as a function of time. Known faulty channels identified by data quality checks~\cite{TCAL-2021-01} are excluded from the cell average. The largest degradation is experienced by the E4 cells covering the pseudorapidity ranges of $1.4<|\eta|<1.6$. The response in the E4~cells is reduced by 41\% on average at the end of the Run~2. Figure~\ref{fig:Ecell_laser} shows the mean response variation measured by the laser system for the PMTs connected to the E-cells as a function of time during the entire Run~2. A maximum degradation of 15\% is observed for the E4-cells.

\begin{figure}[t]
\centering
\includegraphics[width=0.6\textwidth]{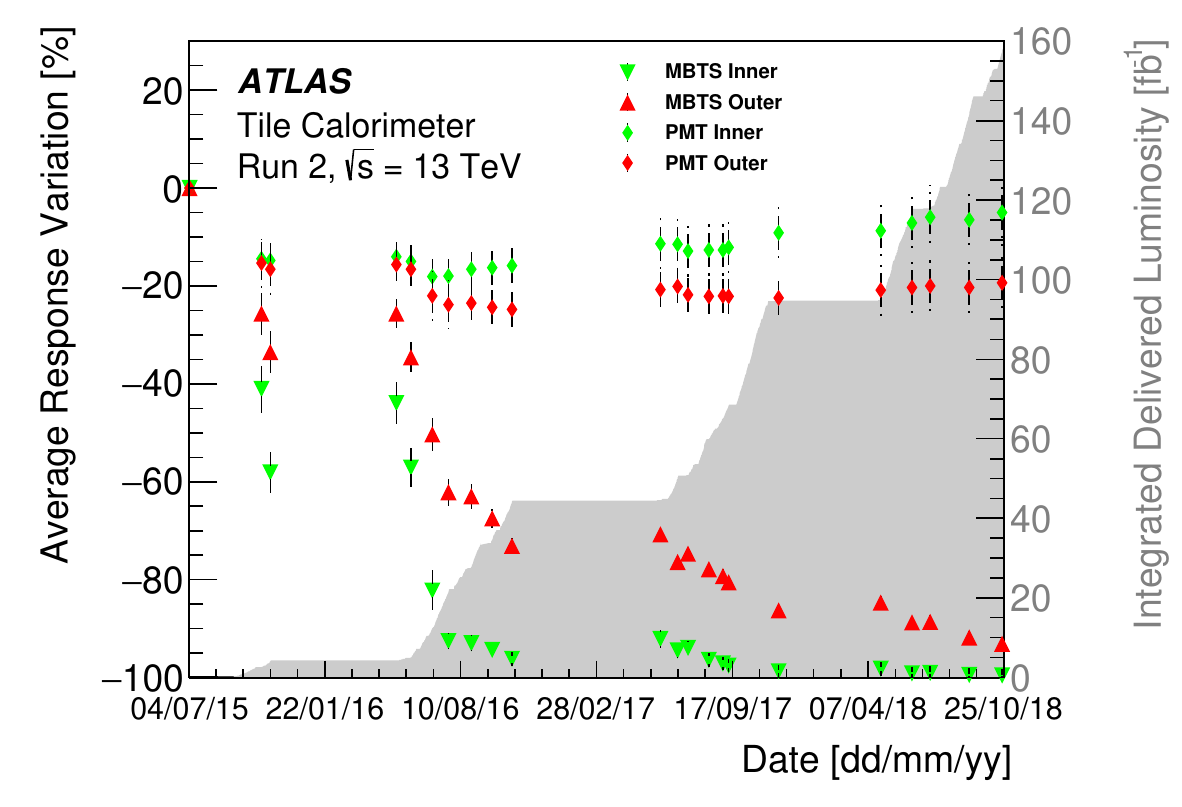}
\caption{The variation of the average response to MB events, $\Delta R_{\mathrm{MB}}$, of the MBTS inner (down triangles) and outer (up triangles) counters as a function of time during Run 2~\cite{TCAL-2021-01}. Response to the MB events is normalised to the track-counting luminosity measured by the inner detector~\cite{DAPR-2021-01}. The circle (diamond) markers show the relative response of the PMTs of the inner (outer) counters to laser pulses, $\Delta R_{\mathrm{Las}}$. The vertical bars represent the RMS of the response over all counters. The LHC delivered luminosity is shown by the shaded area~\cite{DAPR-2021-01}.}
\label{fig:mbts-lumi}
\end{figure}

\subsubsection{Minimum Bias Trigger Scintillators}

The degradation of the MBTS system was determined using the laser calibration and MB data collected during the $pp$ collision runs. The PMTs were operated at a voltage of 500~V. The MBTS response is obtained by normalising the measured MB currents to the ATLAS instantaneous luminosity determined using the track-counting measurements in the inner detector~\cite{DAPR-2021-01}. Figure~\ref{fig:mbts-lumi} shows the average response variation of the MBTS inner and outer counters relative to the first run of Run~2, as a function of time. The relative variations of the average PMT responses measured by the laser system are also shown. The vertical bars represent the RMS of the response over all counters.

At the beginning of Run~2, the mean MB PMT currents were at the level of 15 and 10~nA/10$^{30}$~s$^{-1}$cm$^{-2}$ for the MBTS inner and outer counters, respectively. By the end of 2015 (delivered integrated luminosity of 4.0~\ifb~\cite{DAPR-2021-01}), the MBTS inner (outer) counters have lost almost 55\% (35\%) of their response. This difference is understood by the larger irradiation density of the inner counters, being closer to the beam line. At the beginning of each year's data-taking period, a relative recovery of the response of about 30\% is observed. This happens during the technical stop periods where the readout PMTs partially recover some of their efficiency.

Laser data are used to evaluate the PMT variations alone. The PMTs reading out the inner and outer counters show a rapid decrease of the average response by about 20\% at the beginning of Run~2. This degradation is expected as the PMTs are still receiving high light output from the scintillators at that time, inducing a large anode current and a significant amount of charge being integrated. After this rapid decrease, the responses of the PMTs become more stable and a general up-drift can be noticed. This behaviour is understood by the rapid degradation of the MBTS scintillators/fibres which results in much less light received by the PMTs. Consequently, the integrated PMT anode currents decrease causing the down-drift to cease and eventually to reverse its course.


%

%
\FloatBarrier
\section{Radiation Environment}
\label{sec:radiation}

\begin{figure}[t]
\centering
\includegraphics[width=1.\textwidth]{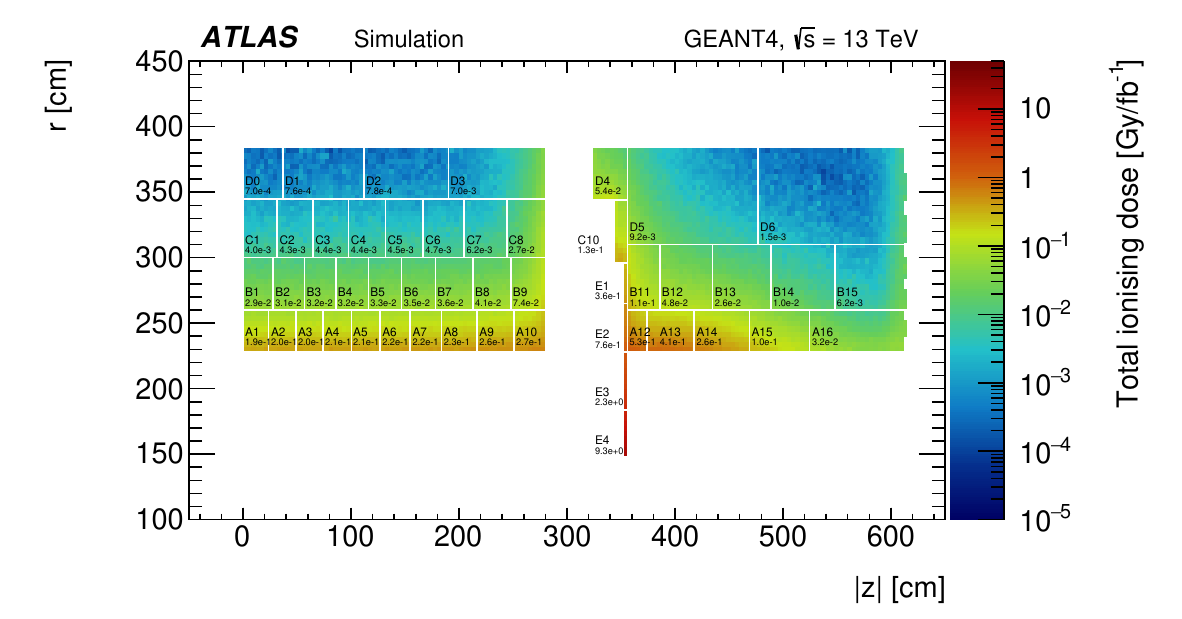}
\caption{\label{fig:Tile_TID_prelim}
Simulated ionisation dose deposited in the scintillator tiles of the cells and in the gap/crack scintillators in $4\times 4$~cm$^2$ bins in $r\times z$. The average dose deposit per cell in Gy per \ifb is also displayed. The study was performed using 50~000 inelastic $pp$ collisions at $\sqrt{s}=13$~\TeV\ generated with \textsc{Pythia~8}. The response of the detector was obtained using the simulation program \textsc{Geant4}. The results are normalised to a cross-section of $\sigma_{\mathrm{inel}}$ = 78.42~mb and an integrated luminosity of 1~fb$^{-1}$~\cite{DoseSimulation,TCAL-2021-01}.}
\end{figure}

During LHC operation, the $pp$ collisions at the interaction point produce a radiation field encompassing the ATLAS detector and the LHC ring~\cite{GENR-2019-02}. Around 50\% of the secondary particles, which are generated when primary particles from the collisions interact with the detector's passive materials or through their decays, are emitted within $|\eta|<3$, in the acceptance region of the central calorimeters, but deposit only about 1\% of the total collision energy there. Particles emitted to the forward region, $3<|\eta|<5$, carry around 5\% of the total collision energy. The remainder of the energy is dissipated in the LHC machine elements and the ATLAS experimental cavern.

Despite dedicated shielding for radiation protection, radiation damage in the sensors, the electronics and power supplies of the ATLAS detector, is an issue, continuously monitored during operation and analysed in the perspective of future upgrades. The radiation environment, composed of several types of particles, is also studied in detailed simulation and monitored in the detector. The ATLAS detector is equipped with 14 RadMon sensors with an accuracy of 20\% to measure the radiation levels in strategic points of the detector~\cite{GENR-2019-02}.

The radiation damage for the TileCal scintillating materials and WLS fibres is a consequence of the energy deposited by ionising processes per unit of mass of the material - total ionising dose (TID). The damage manifests as a reduction in the light response of the scintillator and WLS fibre set as presented in Section~\ref{sec:ageing}. During Run 2, the radiation environment was studied with a \GEANT simulation of a detailed three-dimensional model of the ATLAS detector~\cite{DoseSimulation}. The TID deposited in the TileCal scintillators was simulated for a map of $4\times4$~cm$^2$ in the $(r,z)$-plane and is averaged over $\phi$ even if the underlying geometry is not perfectly $\phi$-symmetric. Mirror-symmetry relative to $z$= 0 is also assumed.

The simulation results are presented in Figure~\ref{fig:Tile_TID_prelim}, showing the TID deposited in the TileCal scintillating material per \ifb of $pp$ collisions in Run~2. The average dose deposit per cell is also displayed. Inner cells, closer to the interaction point, suffer from a harsher radiation environment, with ionising doses on the A12 and A13 cells at the end of Run 2 (integrated luminosity of 157.4~\ifb~\cite{DAPR-2021-01}) of around 85 and 65~Gy, respectively. The TID per \ifb map also shows the large spread of dose within the cells, ranging between 30 and 50\%. This is due to the large cell volumes combined with the large dose gradient. The average dose rates to which cells are subjected are approximated by the product between the simulated cell's average TID in Gy/\ifb and the average instantaneous luminosity of $pp$ collisions. The values are in the order of $10^{-7}$--$10^{-5}$~Gy/s.

Simulations of the expected radiation environment in Run~3 and the HL-LHC were performed using the respective detector geometry description and expected $pp$ run conditions. In particular, upgrades in the inner tracking system and the LAr calorimeter imply a difference in the material in front of the scintillators of the TileCal which affects the TID in scintillators. The resulting dose per \ifb conversion factors are very similar between runs. However, due to the expected increase of instantaneous luminosity in the HL-LHC phase, dose rates are expected to increase proportionally in that detector operation period. More detailed plans for the luminosity profile foresee a staged increase of the peak luminosity with three main operation points: 4, 5 and $7\times 10^{34}$~cm$^{-2}$s$^{-1}$.

From the simulation, the accumulated dose for A12 and A13 is of the order of 250~Gy by the end of Run~3,
considering a total integrated luminosity of 157.4~\ifb in Run~2 and assuming 375~\ifb in Run~3.
For the end of HL-LHC, the total dose obtained from simulation for these cells is around 1500~Gy, assuming a cumulated integrated luminosity of 4000~\ifb and a flat instantaneous luminosity profile with a factor seven increase relative to Run~2 and Run~3 average conditions.

%
%

%

%
%
%
%


%

%
\FloatBarrier
\section{Response degradation of Scintillators and WLS fibres with Dose exposure}
\label{sec:ageing}

The response of the TileCal scintillators and fibres is studied by interplaying information from the different calibration systems. The average light output relative to the beginning of Run~2, $I/I_0$, is defined as

\begin{equation} \label{eq:I/I0}
\centering
I/I_0 = \frac{R_{\mathrm{Cs(MB)}}}{R_{\mathrm{Las}}}
\end{equation}

where $R_{\mathrm{Cs(MB)}}$ and $R_{\mathrm{Las}}$ are the average cell relative response to the caesium source (MB integrated currents) and laser pulses, as detailed in Section~\ref{sec:calibration}. By factoring out the PMT response contribution evaluated with laser from the whole cell response probed with the $^{137}$Cs source or MB integrated currents, the light output of the optical components is isolated and monitored across Run~2.

For the standard calorimeter cells, the $I/I_0$ is evaluated at the cell level and for individual scintillator tiles within the cell (Section~\ref{sec:ageingCells}).
The degradation of the highly exposed gap/crack cells and MBTS system is also measured (Sections~\ref{sec:ageingEcells} and~\ref{sec:ageingMBTS}).
Since the standard calorimeter components are not replaceable for future runs, their data are modelled and extrapolated to future dose exposure conditions (Section~\ref{sec:ageingModel}).

\FloatBarrier
\subsection{Standard Calorimeter Scintillators}
\label{sec:ageingCells}

\subsubsection{Average cell degradation}
\label{average_cell_ageing}

The relative light yield for each TileCal cell is measured using data from the laser system and the Cs scans in Run~2, according to Eq.~(\ref{eq:I/I0}). The quantity is averaged per cell type over the different modules and detector sides. Particular cells with a geometry different from the regular cell map are excluded from the cell average since their radiation conditions and expected degradation profile are not the same as the ones with typical geometry. The measurement is affected by the precision of the laser and Cs systems, which is around 0.5\% for each system~\cite{Cesium_2020,LaserRun2}. These uncertainties are assumed to be independent resulting in about 1\% systematic uncertainty on the $I/I_0$ determinations.

\begin{figure}[t]
\centering
\subfloat[]{\includegraphics[width=.48\linewidth]{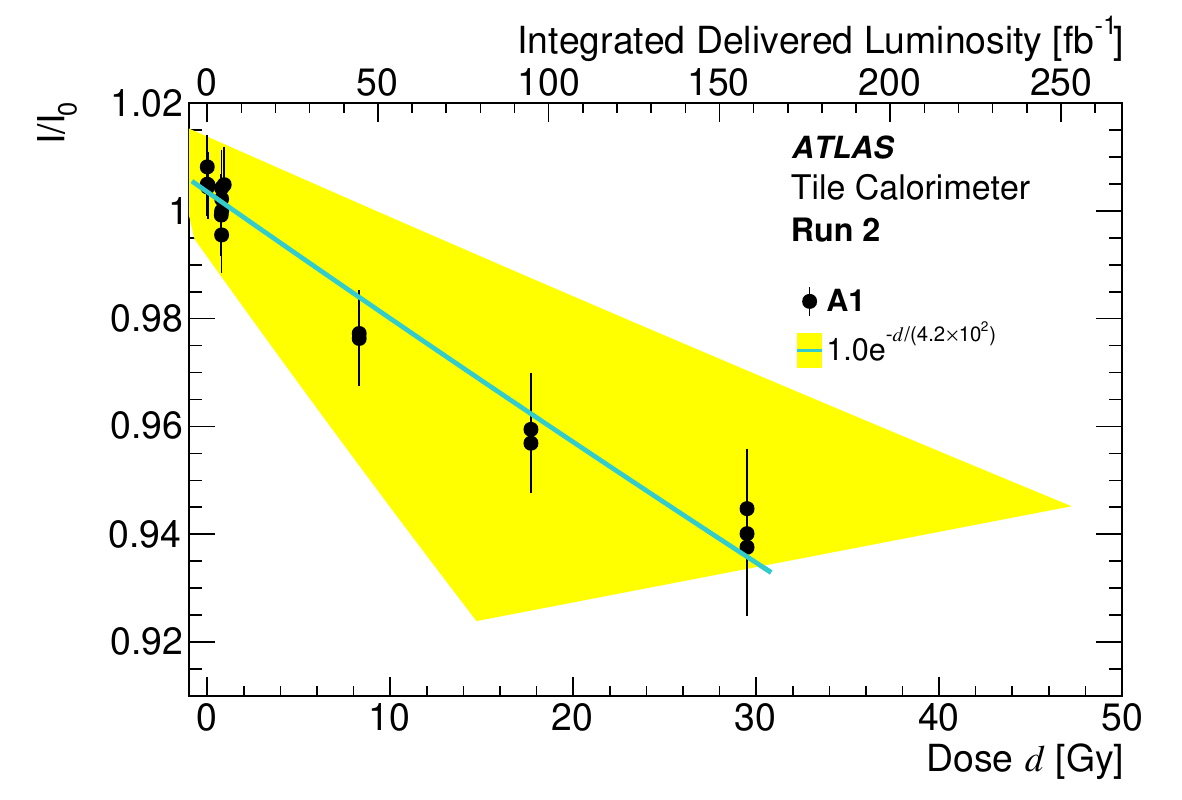}}
\hfill
\subfloat[]{\includegraphics[width=.48\linewidth]{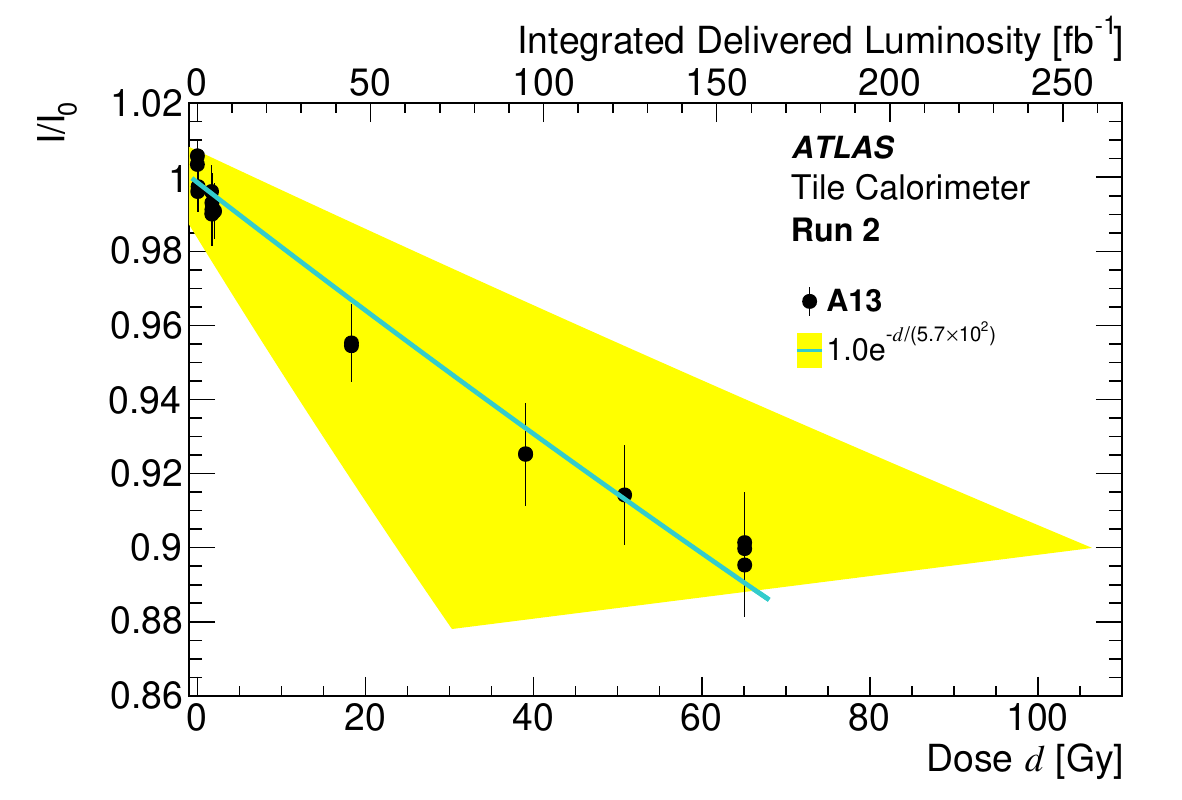}}
\caption{Average relative light yield ($I/I_0$) measurement for (a) A1 and (b) A13 cells as a function of simulated average dose at the time of the LHC Run~2 Cs scans used in the study. The vertical bars represent the RMS of the ($I/I0$) measurement for the different TileCal modules. The line corresponds to the fit of the function in Eq.~(\ref{eq:II0model}) to the data. The yellow region represents the total uncertainty on the fit including the RMS of the dose distribution within the cell and systematic uncertainties on $I/I_0$ due to the intrinsic precision of the caesium and laser measurements.}
\label{fig:ageingVSdose}
\end{figure}

Figure~\ref{fig:ageingVSdose} shows the average $I/I_0$ as a function of the simulated average dose for A1 and A13 cells. A reduction of the light output is observed with the dose deposit, attaining around 6 (10)~\% for A1 (A13) at 30 (65)~Gy. The last A1 (A13) measurements at the same dose of 30 (65)~Gy were obtained from the Cs scans taken after the end of the LHC Run~2 operation, in early 2019, where the detector was no longer subjected to irradiation, except for possible sub-dominant contributions from activation processes. Still, the measurements are compatible within uncertainties.

The degradation of a plastic scintillator, measured for TileCal cells according to Eq.~(\ref{eq:I/I0}), is typically described by having an exponential dependence with the dose $d$~\cite{Zorn1992}, as:

\begin{equation}
I/I_0=p_0e^{-d/p_1}
\label{eq:II0model}
\end{equation}

where $p_0$ is the relative light output before irradiations and $p_1$ is the dose value for which $I/I_0$ is reduced to a fraction $1/e$, characterising how fast the light output degrades with increasing dose exposure. Since a TileCal cell is a collection of scintillators with approximate irradiation conditions, its behaviour is approximate to a single exponential model, and the $p_0$ and $p_1$ parameters are obtained for each cell by minimising the $\chi^2$ of the fit to data as presented in Figure~\ref{fig:ageingVSdose}. The obtained degradation rate is different between the two cells. This is further discussed in Section~\ref{sec:ageingModel}, where the model is also extrapolated to future runs. In Figure~\ref{fig:ageingVSdose}, the yellow region represents the impact of the $I/I_0$ uncertainty (around 1\%) and the RMS of the dose distribution within the cell (around 50\%) on the degradation model. These upper/lower limits result from fitting data points obtained by simultaneously varying up/down the $I/I_0$ measurements by their uncertainty and the doses in the cell by their RMS. Although the dose RMS is not an uncertainty on the average cell dose, a conservative approach incorporating the RMS variation as a systematic error is adopted to determine the model limits for the extreme cases corresponding to the most/least exposed tiles in the cells.

\begin{figure}[t]
\centering
\includegraphics[width=1.\textwidth]{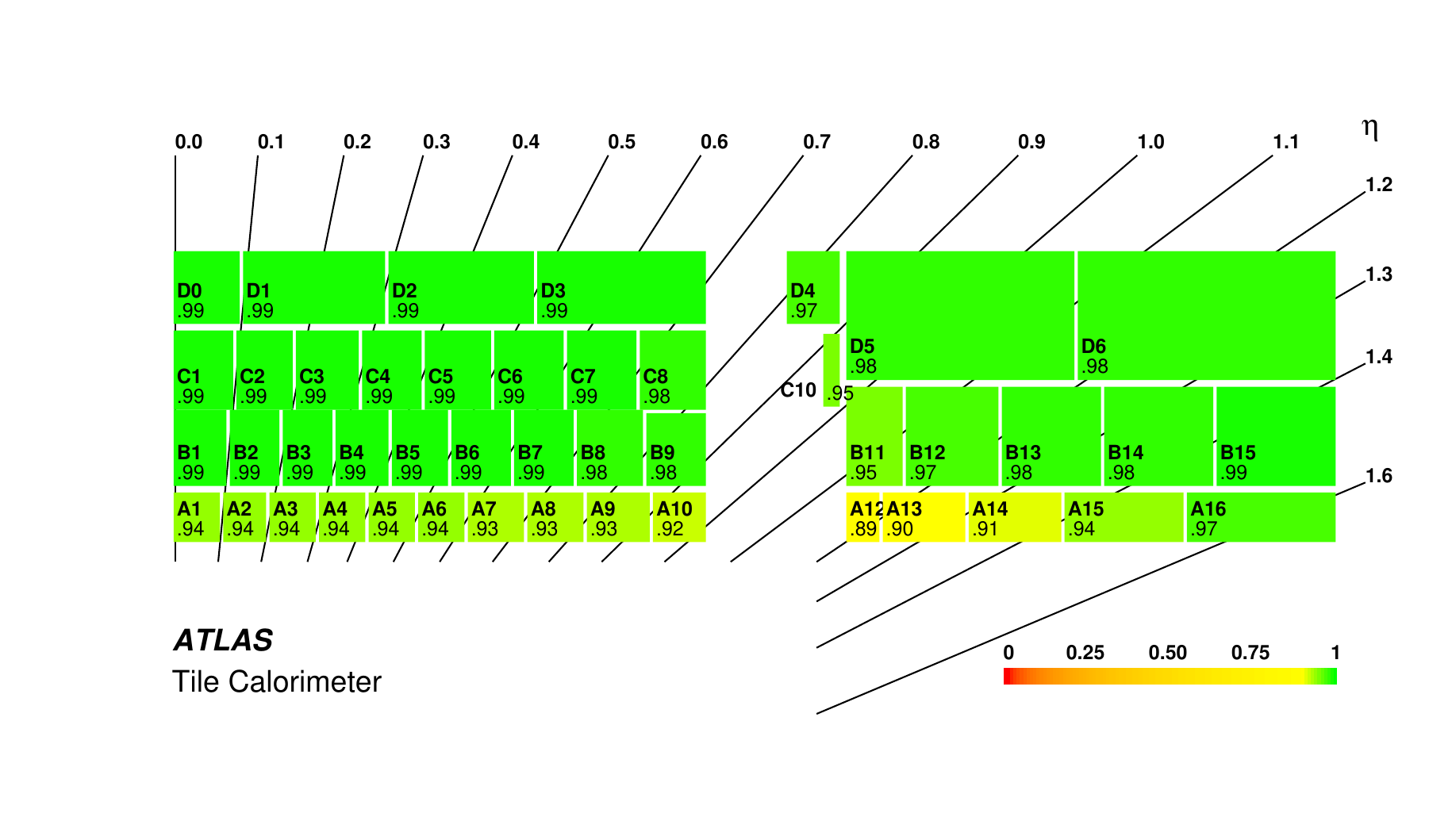}
\caption{\label{fig:II0mapRun2} The measured relative light yield $I/I_0$ as defined in Eq.~(\ref{eq:I/I0}) for the TileCal barrel cells at the end of Run~2~\cite{TCAL-2021-01}. The very correlated uncertainties on the values of each cell are of the order of 1\%, resulting from the quadratic sum of the intrinsic precision of the caesium and laser systems.}
\end{figure}

The average relative light yield measured at the end of Run~2 is shown in Figure~\ref{fig:II0mapRun2} for all the TileCal cells in the barrels.
For around 35\% of the cells, the degradation is about 1\% and the current precision of the measurement does not allow to conclude on their damage. These correspond also to the regions of the detector instrumented with the different types of polystyrene, PSM115 and BASF165H as detailed in Section~\ref{sec:tilecal-optics}, rendering it impossible to study the radiation hardness of each material type. As expected, the A-layer, B11, B12, C10 and D4 cells with the larger total integrated doses have a larger loss in light yield, of the order of 5\% (doses between 10 and 40~Gy). Up to 11\% degradation is observed for A12 and A13 cells with extreme dose accumulation of 85 and 65~Gy, respectively. Still, the detector shows negligible radiation damage overall and was ready for the Run~3 data-taking campaign with no performance deterioration attributable to diminished optics response.

\subsubsection{Single tile-row degradation and cell response uniformity}

The signal amplitudes from each tile in a cell are measured with the Cs calibration data by applying the amplitude method, as described in Section~\ref{subsec:caesium_calibration}. These are used to determine the cell row degradation as a function of time, together with the RMS/mean signal of the tiles within a cell to quantify the cell response uniformity.

The fit performed to determine the tile signal amplitudes does not converge for all cells, which constitutes an additional reason to employ the integral method for calibration. This happens due to known issues related to the $^{137}$Cs scan, such as local instabilities of the source capsule velocity which distort the shape of the response sequence shown in Figure~\ref{fig:example_cs_row_resp}, preventing the correct location of the signal from each tile. The C10 cells, with only five tile/steel plate periods per row and without end plates, also fail to constrain the fit parameters due to insufficient data points~\cite{Cesium_2020}. In addition, damages in the readout system, either in the light readout by WLS fibres or in the electronics cards that process the signal, can prevent the correct determination of signal amplitudes. Cells for which a fit does not converge are removed from further analysis. These amount to around 20-30\% of all the cells analysed, except for D5 and D6, for which this fraction is 50--60\%.

\begin{figure}[t]
\centering
\subfloat[]{\includegraphics[width=0.49\textwidth]{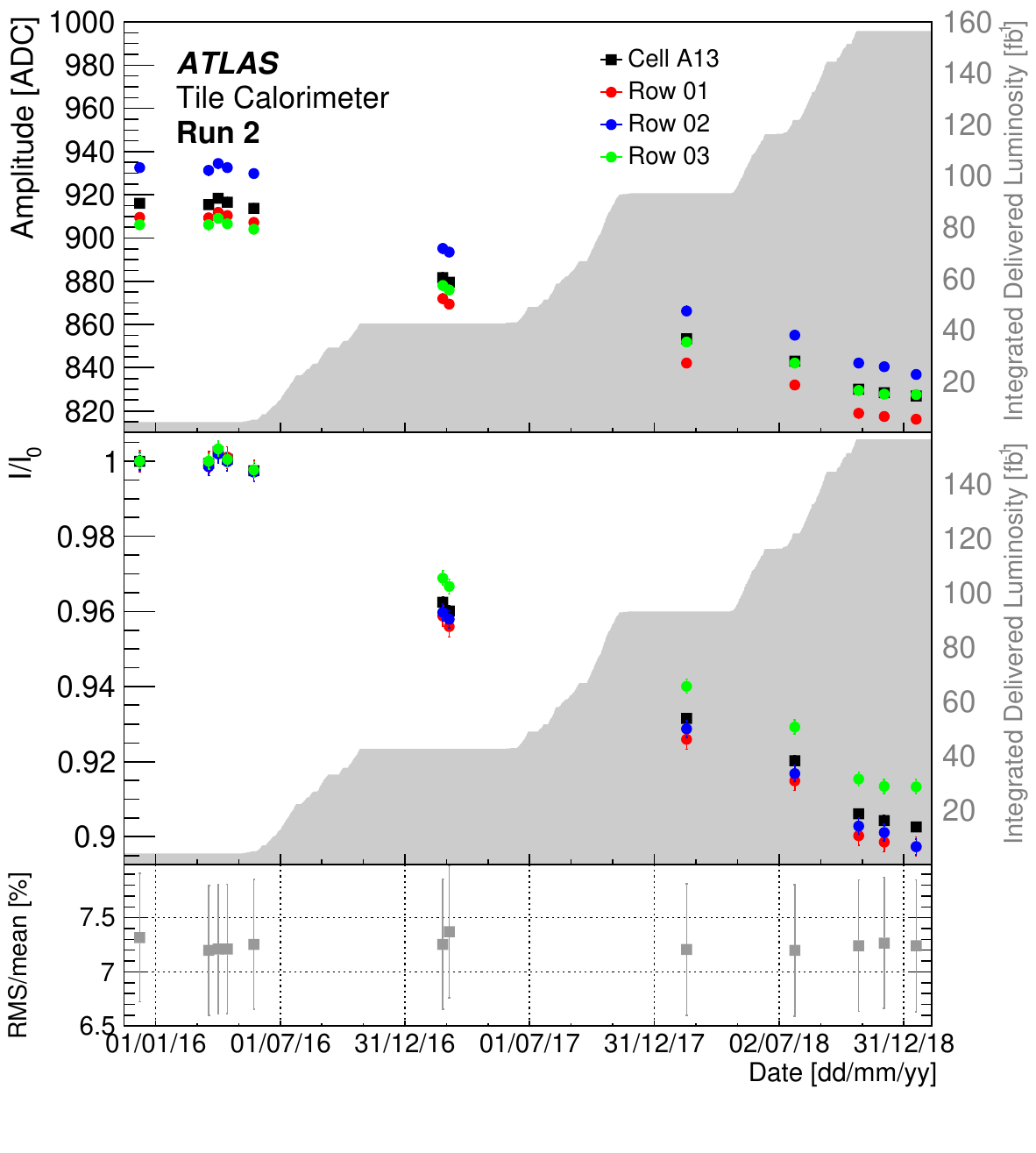}\label{fig:amp__amp_var__RMS_A13}}
\hfill
\subfloat[]{\includegraphics[width=0.49\textwidth]{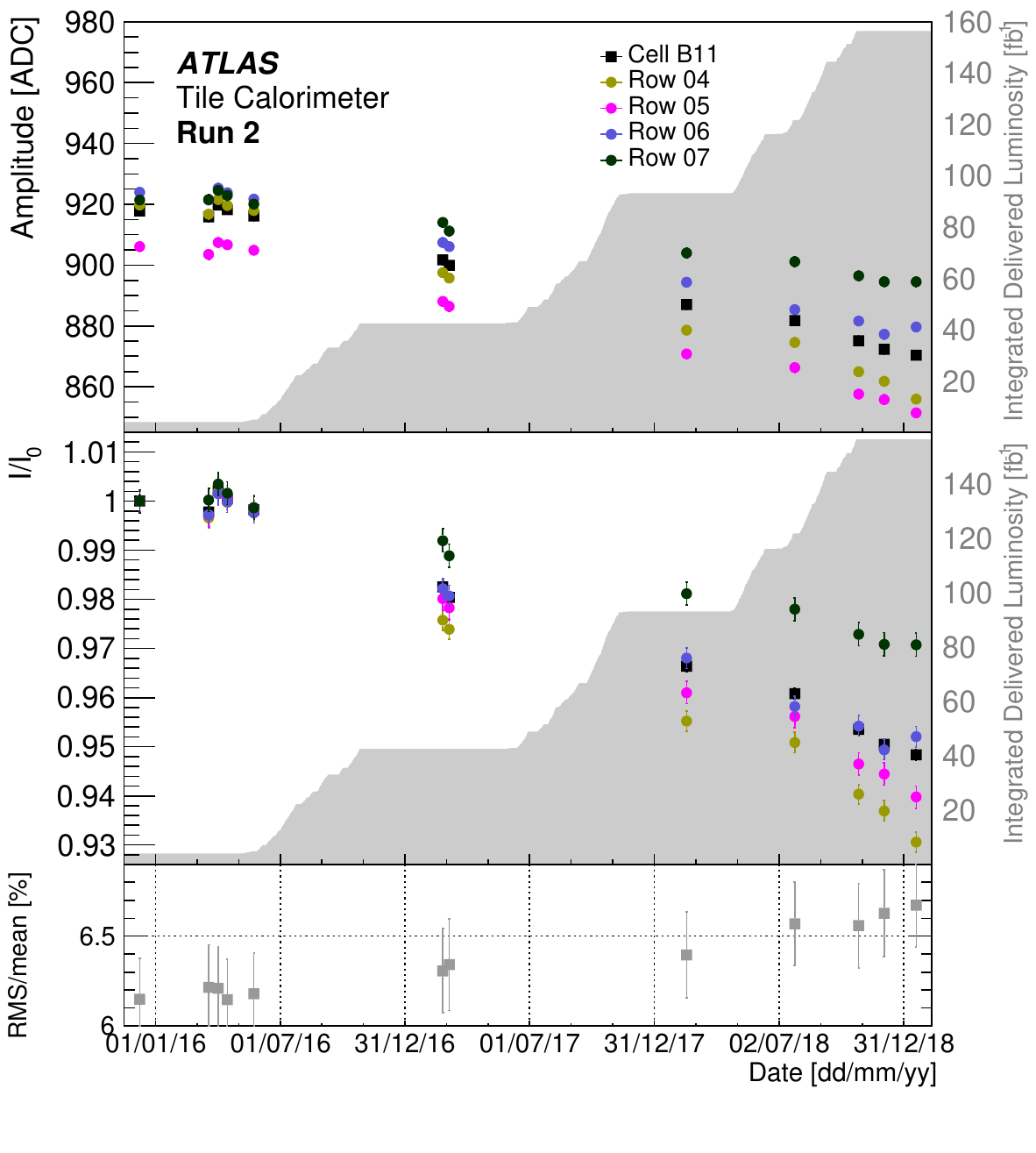}\label{fig:amp__amp_var__RMS_B11}}
\caption{Average amplitude $A$ in ADC counts (upper pad) and $I/I_0$ (middle pad) of the tile rows from (a) A13 and (b) B11 cells as a function of time. The average amplitudes $A$ and $I/I0$ are also displayed for each cell. The bottom pad shows the average RMS/mean amplitude of the cell tiles for (a) A13 and (b) B11 cells as a function of time. Cells for which the amplitude fit did not converge are removed from analysis (around 25\% of the A13 cells and 20\% of the B11 cells). The LHC integrated delivered luminosity is shown in the upper and middle pads by the shaded area~\cite{DAPR-2021-01}. The tile signal amplitudes are calibrated with laser data to factor out the effect of PMT response fluctuations in the amplitude measurement, according to $R_{\mathrm{Las}}$. The vertical bars represent the RMS of the distributions.}
\label{fig:amp__amp_var__RMS}
\end{figure}

The average signal amplitude and $I/I_0$ of the three tile rows from the A13 cells are displayed in Figure~\ref{fig:amp__amp_var__RMS_A13} as a function of time in the upper and middle pads, respectively. The average cell $I/I_0$ is also presented and shows very good agreement with the results using the caesium integral method presented in Figure~\ref{fig:ageingVSdose}.
The row responses decrease in time owing to the radiation damage of the scintillating tiles and WLS fibres, and at a different rate depending on their radiation exposure profile. Comparing the different rows in A13 at the end of Run~2, a maximum light output degradation of 10.5\% is observed for the innermost one (row 1), while the most shielded of the three (row~3) presents the minimum degradation of 8.5\%. Although small, these differences in ageing rate may impact the cell response uniformity, measured as the RMS/mean signal of the constituent tiles. The average RMS/mean of the A13 cells is shown in the bottom plot of Figure~\ref{fig:amp__amp_var__RMS_A13} as a function of time. There is a large RMS associated with the distribution of RMS/mean for the A13 cells of the different TileCal modules (vertical bars) but it is still useful to analyse the trend of the average RMS/mean with time. It is observed that the RMS/mean value is mostly constant during the whole of Run~2, resulting from the initial row amplitudes (upper pad) and their re-ordering in time due to degradation at different rates. %

Figure~\ref{fig:amp__amp_var__RMS_B11} presents the same measurements of average signal amplitude, $I/I_0$ and response RMS/mean for B11 cells, composed of four tile rows, as a function of time. The total radiation damage in B11 rows at the end of Run~2 is between 3 and 6\% and lower than in A13, with the corresponding cell-level measurement being compatible with the one obtained from the integral method presented in Figure~\ref{fig:II0mapRun2}. However, the differences in the B11 ageing rates are more noticeable than in A13 owing to the larger differences in the radiation environment. For this cell, the average RMS/mean increases towards the end of Run~2, after the first stable period resulting from the combination between particular starting conditions of the tiles's responses and their degradation rates. A maximum of around 7\% non-uniformity is observed.

This study was performed for the whole EBA partition, excluding the cells with special geometry mentioned previously. The average RMS/mean cell uniformity at the beginning of Run~2 was between $\sim$6\% for B13 cells and $\sim$9\% for D4 cells. Although a response degradation is observed for the rows of the most exposed cells, the change in the average response uniformity during Run~2 is not significant for most of the cells. As discussed, this is attributed to the particular initial responses of the scintillators in the cells. The study shows that once the absolute signal of the tiles degrading faster gets smaller than the signal of tiles degrading slower, the non-uniformity increases depending on the amount of radiation damage to the cell. At the end of Run~2, the D4 cell presented the maximum signal RMS/mean value of $\sim$9\% and all the EBA cells studied responded with a non-uniformity below the 10\% specifications of TileCal~\cite{ATLAS-TDR-03}.

\subsection{Gap/Crack Scintillators}
\label{sec:ageingEcells}

The E-cells face the largest irradiation conditions for being placed in the gap/crack region of the detector, less shielded from ionising particle exposure as shown in Figures~\ref{fig:ATLAScalorimeters} and~\ref{fig:tilecal_cell_map}. According to the simulations presented in Section~\ref{sec:radiation}, the gap scintillators accumulated a total dose of around 60 and 120~Gy during the LHC Run~2 (E1 and E2 cells, respectively), the same dose magnitude as the most exposed standard calorimeter cells in the barrel (A12 and A13). The crack cells, however, experienced the most extreme particle fluence, accumulating around 400 and 1500~Gy dose during the same period (E3 and E4, respectively).

The response of the E-cells was measured during Run~2 using the minimum bias and laser data according to Eq.~(\ref{eq:I/I0}). The light output loss of E1 and E2 cells is up to 12\%, as shown in Figure~\ref{fig:II0-Ecells}. The figure also shows the wide spread of the measurement across the TileCal modules, quantified by the RMS represented as vertical bars. This can be associated to different degradation rates of the scintillators, which in principle follows a normal law, and to residual $\phi$ asymmetry on the detector geometry and particle fluence. For E3 and E4 cells, the cumulative loss during the LHC Run~2 is around 20\% and 30\%, respectively. Given the large size and particular geometry of the single scintillators instrumenting these cells, the radiation environment is significantly variable within the scintillator body. This results in a large dose non-uniformity restricting the degradation modelling and accurate extrapolations. Due to the additional expected dose from the 375~\ifb of Run~3 collisions and the current light output lost by E3 and E4 cells, these counters were replaced in 2019--2020 by 6~mm thick EJ-208 scintillators from Eljen, designed with an improved radiation hardness~\cite{GENR-2019-02}. In this upgrade, the pseudorapidity coverage of E-cells was extended to $1.2<|\eta|<1.72$.

\begin{figure}[t]
\centering
\includegraphics[width=0.6\textwidth]{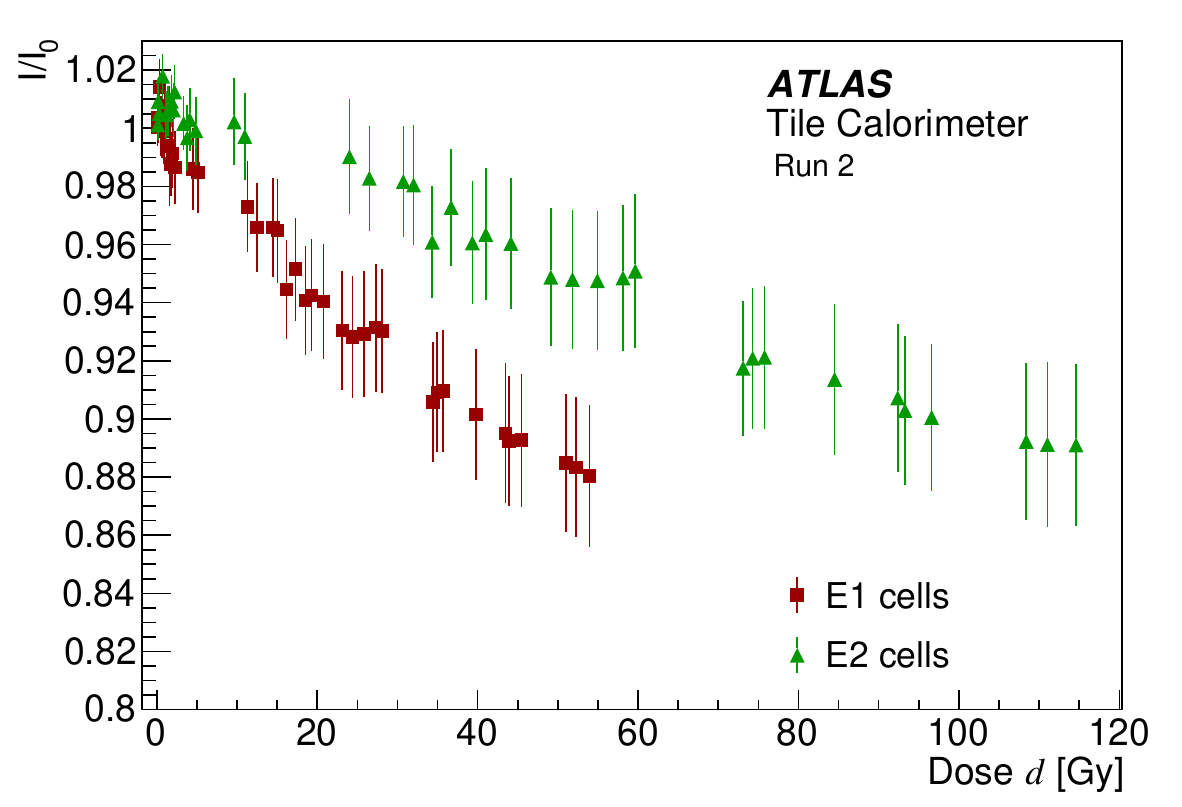}
\caption{Average relative light yield ($I/I_0$) of the E1 (squares) and E2 (triangles) cell as a function of average simulated dose during the LHC Run~2. The vertical bars correspond to the RMS of the cells response distributions.}
\label{fig:II0-Ecells}
\end{figure}

\subsection{Minimum Bias Trigger Scintillators}
\label{sec:ageingMBTS}

Due to the high irradiation dose, scintillator, fibre and PMT performances degraded throughout Run~2
and large MBTS signal reductions were observed. The MBTS signal reductions follow a decaying curve as a function of the irradiation dose, according to the typical degradation of plastic scintillators~\cite{Zorn1992}.

The average total ionising doses, estimated using the method described in Section~\ref{sec:radiation}, are $0.62\times 10^3$~Gy/fb$^{-1}$ for the inner counters and $0.83\times 10^2$~Gy/fb$^{-1}$ for the outer counters. %
For the same integrated luminosity, the inner counters accumulate an order of magnitude more dose than the outer counters.

To measure the light yield degradation of the MBTS scintillators and fibres, PMT variations are factored out from the minimum bias data, as expressed in Eq.~(\ref{eq:I/I0}).
The results as a function of the total ionising dose are shown in Figure~\ref{fig:mbts-dose}. The decline follows an exponential decay curve until around 15~kGy with a similar degradation rate for the inner and outer counters. With the increase of the doses, the inner counters continue to decrease exponentially with a slowing rate presumably caused by saturation of the depleting effects in the scintillating material.

Due to the severe damage in Run~2, the MBTS counters were replaced for Run~3~\cite{GENR-2019-02}. Polystyrene-based scintillators were installed for both the inner and outer segments but the innermost counters are doped with PTP (primary dopant) and BBQ (secondary dopant). For these, the scintillation is in the green region of the spectrum, which is less affected by absorption in the colour centres produced in radiation exposure. The outer segment used the standard PTP and POPOP dopants, producing blue scintillator light.

\begin{figure}[t]
\centering
\includegraphics[width=0.6\textwidth]{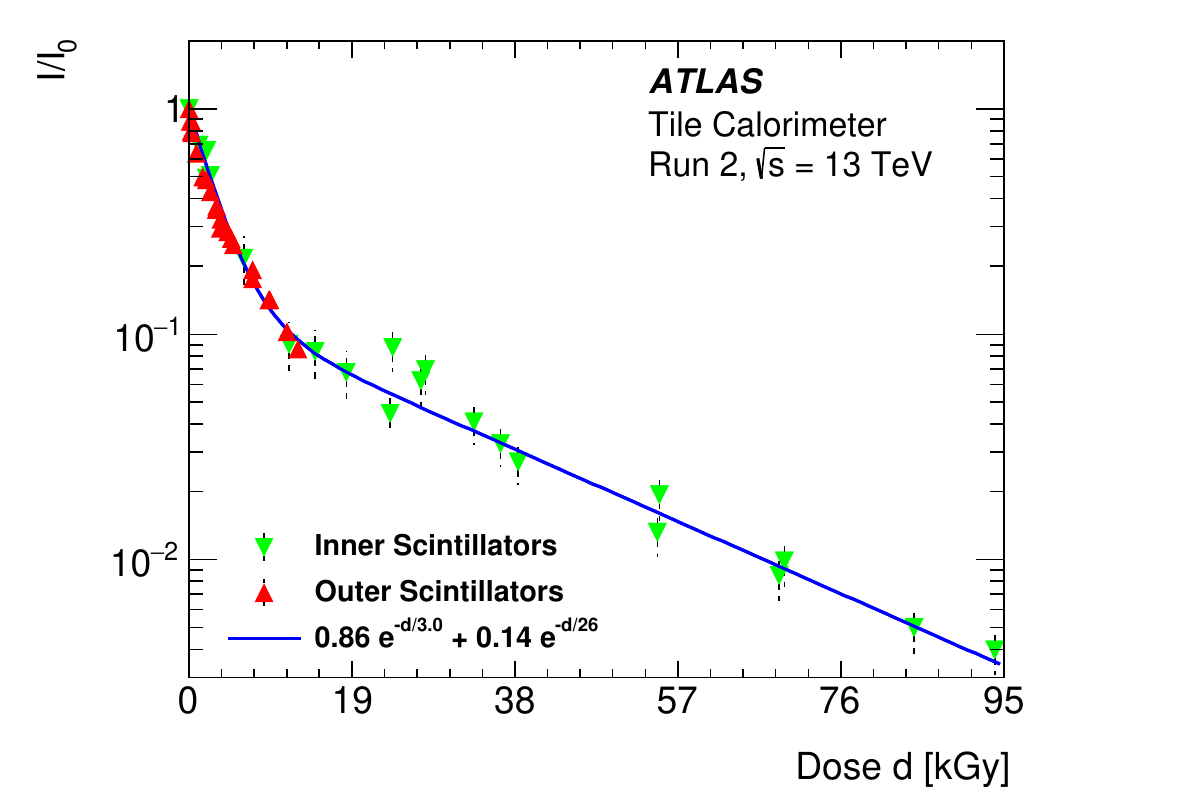}
\caption{Average relative light yield ($I/I_0$) of scintillators and fibres of the inner (down triangles) and outer (up triangles) counters as a function of total ionising dose during Run~2~\cite{TCAL-2021-01}. The values are the averages of the corresponding determinations obtained for the inner and outer counters. The vertical bars correspond to the RMS of the counter response distributions. The curve obtained by fitting the inner MBTS data points is drawn (solid line) and the corresponding function is displayed in the legend.}
\label{fig:mbts-dose}
\end{figure}

\subsection{Dose Rate effects and Extrapolation to future conditions}
\label{sec:ageingModel}

The HL-LHC phase, targeting 4000~\ifb of total integrated luminosity from $pp$ collisions, will represent additional radiation exposure for the TileCal. In particular, the increase of instantaneous luminosity by a factor of seven will result in higher dose rate conditions~\cite{ZurbanoFernandez:2020cco}. This environment challenges the calorimeter performance due to the progressive fade out of the sensitive material towards the end of the experiment's lifetime. Under these circumstances, modelling the scintillator radiation damage with Run~2 data and extrapolating it to future operation is essential to adopt mitigating strategies beforehand, such as instrumenting the readout of the more damaged cells with larger efficiency photodetectors~\cite{ATLAS-TDR-28}.

\subsubsection{Dose Rate effects and degradation model}

\begin{figure}[t]
\centering
\includegraphics[width=0.7\textwidth]{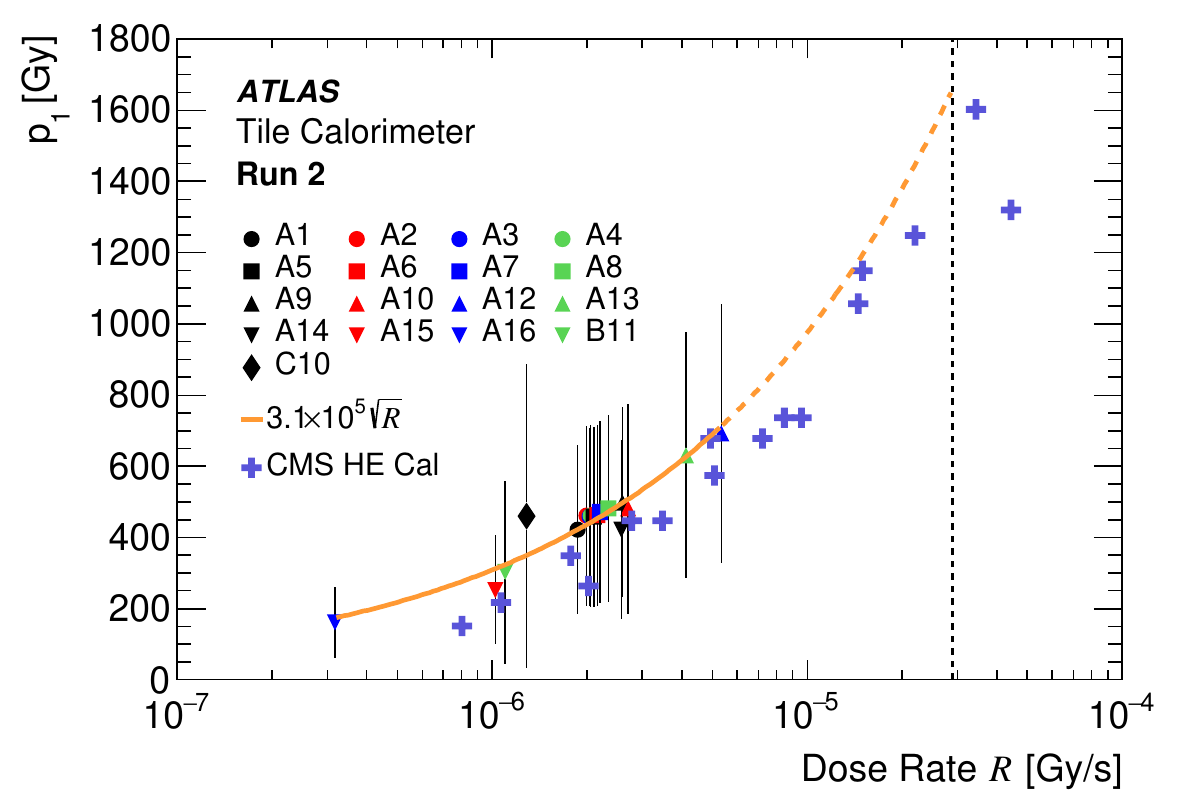}
\caption{\label{fig:dose_rate}
The degradation rate parameter $p_1$ obtained from the simple exponential model in Eq.~(\ref{eq:II0model}) as a function of the average simulated dose rate $R$ for the most exposed cells~\cite{TCAL-2021-01}. Results from a similar study performed using the CMS Hadron Endcap Calorimeter measurements are also displayed (crosses)~\cite{CMSHCAL:2016dvd}. Due to its forward location, the CMS HE Cal detector experiences dose rates higher than the TileCal, accummulating larger doses for the same luminosity. The vertical error bars on the TileCal data points represent the very correlated total uncertainty including the $I/I_0$ measurement uncertainty and the dose spread within the cell volume. The nominal points are fitted with a power law function reported in the legend (continuous curve). This function is extrapolated to the higher dose rate region (dashed curve) expected at the HL-LHC phase and populated by the CMS HE Cal data. The dashed vertical line indicates the expected dose rate of the A13 cells in the HL-LHC.}
\end{figure}

Studies of radiation damage to plastic scintillators installed at the CMS Hadron Endcap Calorimeter (HE Cal) have shown that dose rate conditions may significantly affect the degradation rate concerning dose, with larger dose rates corresponding to smaller degradation rates~\cite{CMSHCAL:2016dvd}. While the effect is nearly irrelevant for the LHC Run~3, where dose rates are similar to the Run~2 ones, extrapolating to the HL-LHC requires the integration of possible dose rate effects in the exponential model described by Eq.~(\ref{eq:II0model}). Figure~\ref{fig:dose_rate} shows the degradation rate parameter $p_1$ obtained by fitting Run~2 data from the most degraded TileCal cells as a function of their average dose rate conditions. As stated in Section~\ref{sec:radiation}, the average dose rates are approximated by the product between the cell average dose in units of Gy/\ifb and the average instantaneous luminosity in $pp$ collisions during Run 2. The results exhibit the same trend as CMS. The CMS HE Cal data are also plotted and the agreement with TileCal measurements is quantitatively good. Yet, there are important differences between the optics instrumentation of the two calorimeters -- CMS HE Cal features polystyrene-based scintillators from Kuraray (SCSN-81)~\cite{CMSHCAL:2016dvd} in the first and seventh HE Cal layers displayed in Figure~\ref{fig:dose_rate}.

To model the dose rate effect, the TileCal data are fitted with a power law function, shown in the legend of Figure~\ref{fig:dose_rate} (solid orange line). This follows a degradation model based on oxygen diffusion, where permanent damage is attributed to opaque colour centres formed by the interaction between oxygen molecules and radicals created during irradiation, and the reformation of chemical bounds enhanced by radical mobility~\cite{TRIMMER199357,BUSJAN199989}. The power law allows extrapolating the degradation rate to the HL-LHC phase. For instance, assuming a flat profile of instantaneous luminosity in the HL-LHC, a factor seven larger than in Run~2 and Run~3, the average dose rate for A13 cells is expected to increase from $4.1\times 10^{-6}$ to $2.9\times 10^{-5}$~Gy/s, and the exponential decay parameter $p_1$ is foreseen to increase from 630~Gy in Run~2 to 1670~Gy in the HL-LHC phase (dashed orange line), corresponding to a decrease of degration rate.

\begin{figure}[t]
\centering
\includegraphics[width=0.7\textwidth]{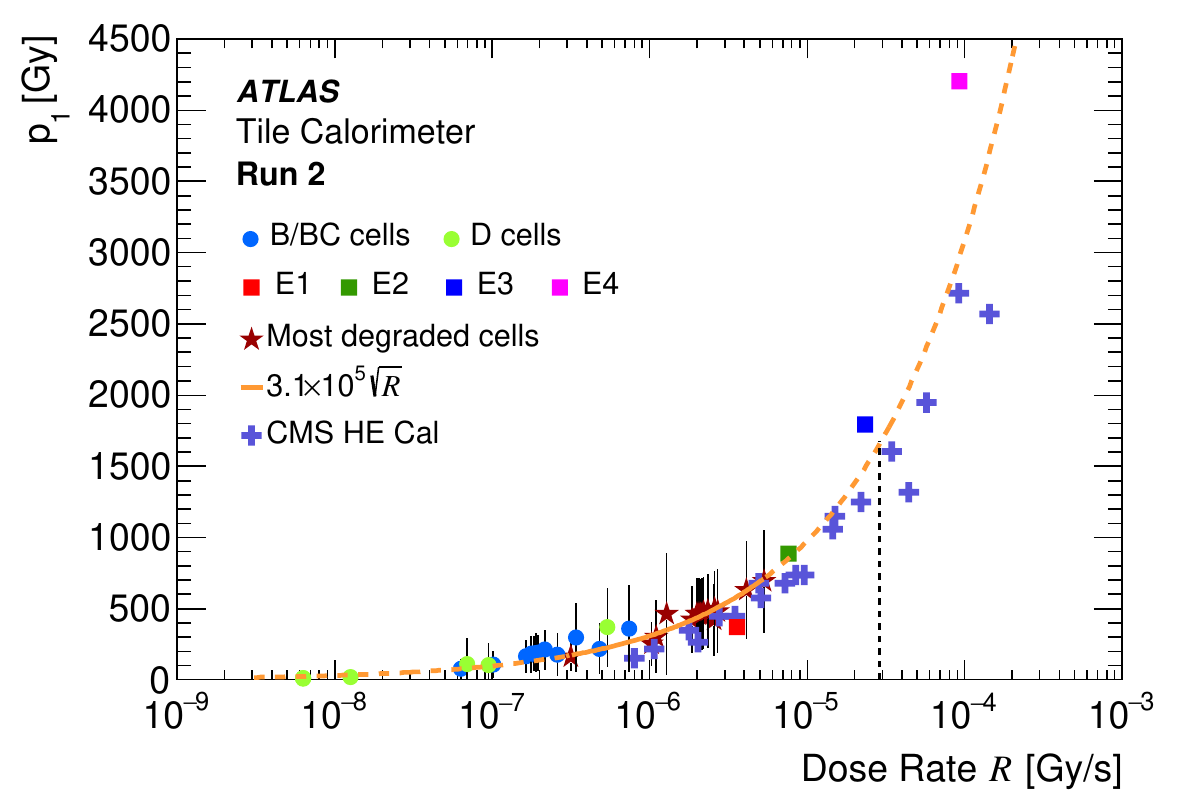}
\caption{\label{fig:dose_rate_consistency}
The degradation rate parameter $p_1$ obtained from the simple exponential model in Eq.~(\ref{eq:II0model}) as a function of the average simulated dose rate $R$ for the most (stars) and least (circles) exposed cells, and E-cells (squares). Results from a similar study performed using the CMS HE Cal measurements are also displayed (crosses)~\cite{CMSHCAL:2016dvd}. Due to its forward location, the CMS HE Cal detector experiences dose rates higher than the TileCal, accummulating larger doses for the same luminosity. The vertical error bars on the TileCal data points from most exposed cells represent the very correlated total uncertainty including the $I/I_0$ measurement uncertainty and the dose spread within the cell volume. The nominal points from these cells are fitted with a power law function reported in the legend (continuous curve). This function is extrapolated to the higher dose rate region (dashed curve) expected at the HL-LHC phase and populated by the CMS HE Cal data. The dashed vertical line indicates the expected dose rate of the A13 cells in the HL-LHC.}
\end{figure}

The consistency of the degradation model is further verified by comparing its extrapolation to lower dose rates with the degradation parameters $p_1$ obtained for the least exposed cells of the TileCal barrels (from B/BC- and D-layers), as shown in Figure~\ref{fig:dose_rate_consistency}. Since the observed light output loss in Run~2 is negligible for these cells, the sensitivity to radiation damage is more weak than for cells in the A-layer. Nevertheless, the agreement between the model and measurements is good. Furthermore, the validity of the model at larger dose rates is similarly probed with data obtained from studying the light response of the more exposed E-cells scintillators, also displayed in Figure~\ref{fig:dose_rate_consistency}. The same qualitative anti-correlation is observed, with larger dose rates associated with smaller degradation rates. The agreement between the model and data is good, but the precision of such quantitative comparisons is limited by the large variation of exposure conditions across the extended geometry of the gap/crack scintillators. The largest difference between the degradation model and data is observed for E4 cells, where the dose rate regime is a few orders of magnitude larger than the HL-LHC extrapolation conditions for the barrel cells.

\subsubsection{Extrapolation to the LHC Run~3 and the HL-LHC phase}

\begin{figure}[t]
\centering
\includegraphics[width=0.7\textwidth]{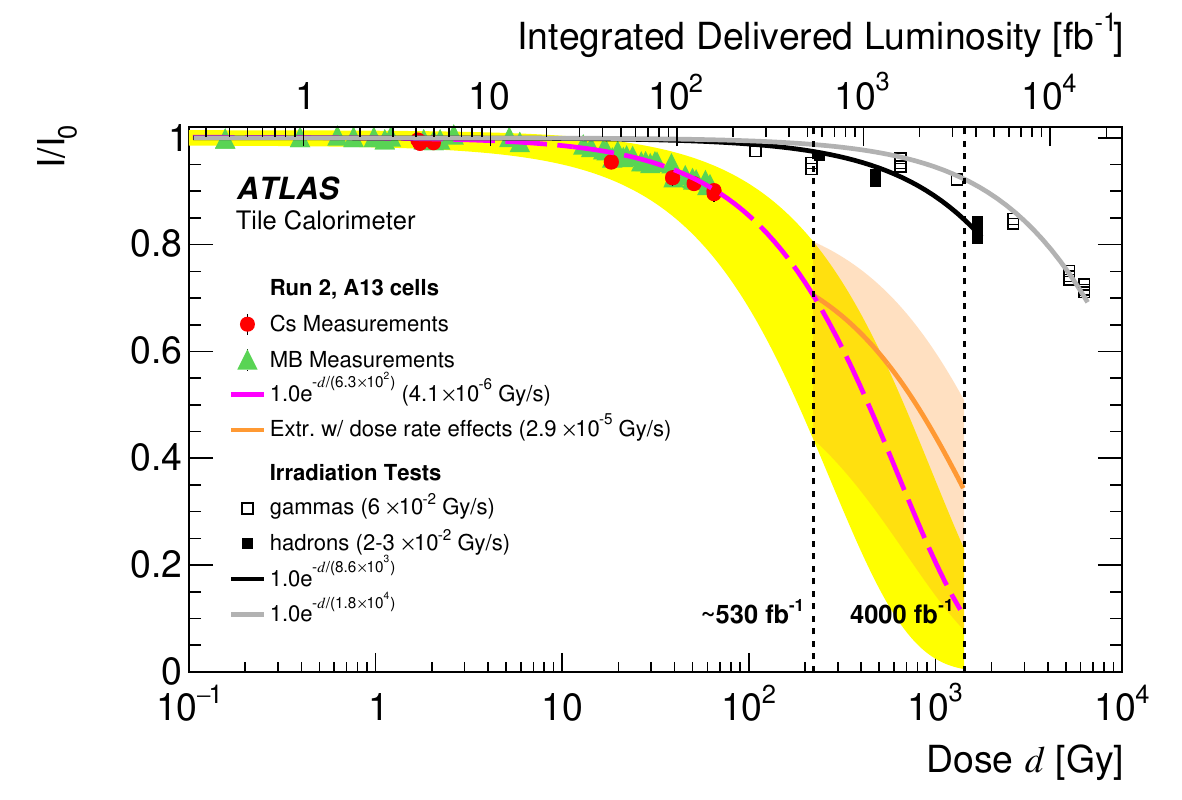}
\caption{\label{fig:II0_A13_log}
Average relative light yield ($I/I_0$) measurements based on the caesium system (dots) and integrated minimum-bias currents (triangles) for A13 cells as a function of average simulated dose $d$ and LHC integrated luminosity~\cite{TCAL-2021-01}. The vertical bars, smaller than the data points, correspond to the RMS of the distribution. The dashed curve corresponds to the fit to the function in Eq.~(\ref{eq:II0model}) to the data. The corresponding average dose rate is shown in brackets. The surrounding opaque region represents the total uncertainty in the fit including the RMS of the dose distribution within the cell and systematic uncertainties in $I/I_0$ due to the intrinsic precision of the caesium, MB and laser measurements. The solid curve represents the expected average $I/I_0$ of the A13 cells in the HL-LHC phase including dose rate effects. The dose rate value in the extrapolation is shown in brackets. The surrounding semi-transparent region is the total uncertainty on this extrapolation, obtained by propagating the uncertainty sources of the study. Results from measurements of bare scintillators performed one month after irradiations made in the laboratory before the detector construction are also shown~\cite{Abdallah:2007cca}. An exponential function is fitted to the data obtained from irradiations with $\gamma$s (open squares) and hadrons (full squares). The corresponding dose rates are shown in the legend. Dashed vertical lines represent the expected dose by the end of the LHC Run~3 ($\sim$530~fb$^{-1}$) and HL-LHC (4000~fb$^{-1}$)~\cite{ZurbanoFernandez:2020cco}.
}
\end{figure}

Figure~\ref{fig:II0_A13_log} shows the average $I/I_0$ for A13 cells as a function of dose, together with the exponential fit to the data as described in Section~\ref{average_cell_ageing} and the data extrapolation to the Run~3 and HL-LHC periods, considering or not dose rate effects in the degadation model (solid {\emph{versus}} dashed line, respectively).
An uncertainty region taking into account the total 1\% systematic uncertainty on the $I/I_0$ measurements and the spread of the dose within the cell is drawn.
This is obtained by fitting the corresponding uncertainty variations from the nominal measurements, simultaneously for both sources, as detailed in Section~\ref{average_cell_ageing}.
As discussed, dose rate effects are significant for the HL-LHC phase. Assuming an increase by a factor of seven in luminosity, the expected degradation at the end of the HL-LHC, after 4000~\ifb, is shifted almost by a factor of two to $34^{+17}_{-26}\%$ residual light yield. The adoption of a more realistic luminosity profile, following the staged increase discussed in Section~\ref{sec:radiation}, only marginally reduces the extrapolated light output since the dominant dose deposit in scintillators arises from the HL-LHC operation at $7\times 10^{34}$~cm$^{-2}$s$^{-1}$ instantaneous luminosity.

While this result takes into account the experimental uncertainties related to the actual measurements, it is important to highlight that the extrapolation model does not take into account any possible scintillator recovery or annealing influences that might result from long periods without LHC collisions. The results in Figure~\ref{fig:tile_irradiation_lab} of Section~\ref{sec:tilecal-optics} from the TileCal scintillator irradiations at detector development are also plotted in Figure~\ref{fig:II0_A13_log} for reference. It is observed that TileCal scintillators present a larger degradation with in-situ conditions, partially attributable to the much lower dose rates faced in the experiment. However, the data are not fully comparable since the laboratory irradiations were done to bare scintillators and the measurements were taken after a month, with significant recovery from the suffered damage.

\begin{table}[t]
\caption{Simulated dose and expected relative light yield $I/I_0$ degradation ($1-I/I_0$ in percentage) of the TileCal cells at the end of the Run~3 and the HL-LHC, assuming a total integraded luminosity of $\sim$~530~\ifb and 4000~\ifb, respectively. Results for the cells of the different layers and for the most exposed cells in each layer are presented. For the HL-LHC runs, the assumed dose rate is seven times higher than in Run~2 and Run~3. The relative uncertainty on the extrapolation to the end of Run~3 is of the order of 5\% and on the extrapolation to the end of the HL-LHC is up to 50\% for the cells with significant expected degradation (larger than 50\%).}
\label{tab:extrapolationR3HL}
\begin{center}
\begin{tabular}{l|c|c|c|c|}
\cline{2-5} & \multicolumn{2}{c|}{End of Run~3}  & \multicolumn{2}{c|}{End of HL-LHC}  \\ \hline
\multicolumn{1}{|l|}{Cell(s)} & \multicolumn{1}{c|}{\begin{tabular}[c]{@{}c@{}}Simulated \\ Dose {[}Gy{]}\end{tabular}} & \begin{tabular}[c]{@{}c@{}}Expected $I/I_0$ \\ degradation {[}\%{]}\end{tabular} & \multicolumn{1}{c|}{\begin{tabular}[c]{@{}c@{}}Simulated \\ Dose {[}Gy{]}\end{tabular}} & \begin{tabular}[c]{@{}c@{}}Expected $I/I_0$ \\ degradation {[}\%{]}\end{tabular} \\ \hline
\multicolumn{1}{|l|}{Typical A-layer}  & $[55,140]$ & $[20,28]$ & $[400,1000]$ & $[48,60]$ \\
\multicolumn{1}{|l|}{A16}              & 18         & $\sim$10  & 140          & $\sim$30 \\
\multicolumn{1}{|l|}{A12, A13}         & 280, 220   & $\sim$30  & 1750, 1400   & 69, 66 \\
\hline
\multicolumn{1}{|l|}{Typical B/BC-layer} & $[4,20]$  & $[5,8]$  & $[30,125]$  & $[17,25]$ \\
\multicolumn{1}{|l|}{B9, B11, B12, C10}  & $[25,70]$ & $[10,15]$ & $[170,470]$ & $[32,44]$ \\
\hline
\multicolumn{1}{|l|}{Typical D-layer} & $[0.4,5]$ & $[3,5]$ & $[3,35]$ & $[7,16]$ \\
\multicolumn{1}{|l|}{D4} & 30 & 12 & 195 & $\sim$30 \\
\hline
\end{tabular}
\end{center}
\end{table}

The analysis presented for the A13 cells, in Figure~\ref{fig:II0_A13_log}, was done for all the TileCal barrel cells and their light output degradation was extrapolated for Run~3 and HL-LHC. The results on the expected average $I/I_0$ determined for the end of Run~3 ($\sim$530~\ifb expected integrated luminosity) and HL-LHC (expected 4000~\ifb) are summarised in Table~\ref{tab:extrapolationR3HL}. In Run~3, it is expected the same degradation profile as in Run~2 and a factor two of additional light output loss. A-layer cells have an expected light yield loss of up to 28\% (doses between 55 and 140~Gy) and up to 33\% (for A12 and A13 at 280 and 221~Gy, respectively). The great majority of the other detector layers are expected to present less than 10\% response loss due to radiation damage at the beginning of the HL-LHC phase. For the same layer, cells located in the EB are expected to be more affected than in the LB. The relative uncertainty on this extrapolation is of the order of 5\%.

\begin{figure}[t]
\centering
\includegraphics[width=\textwidth]{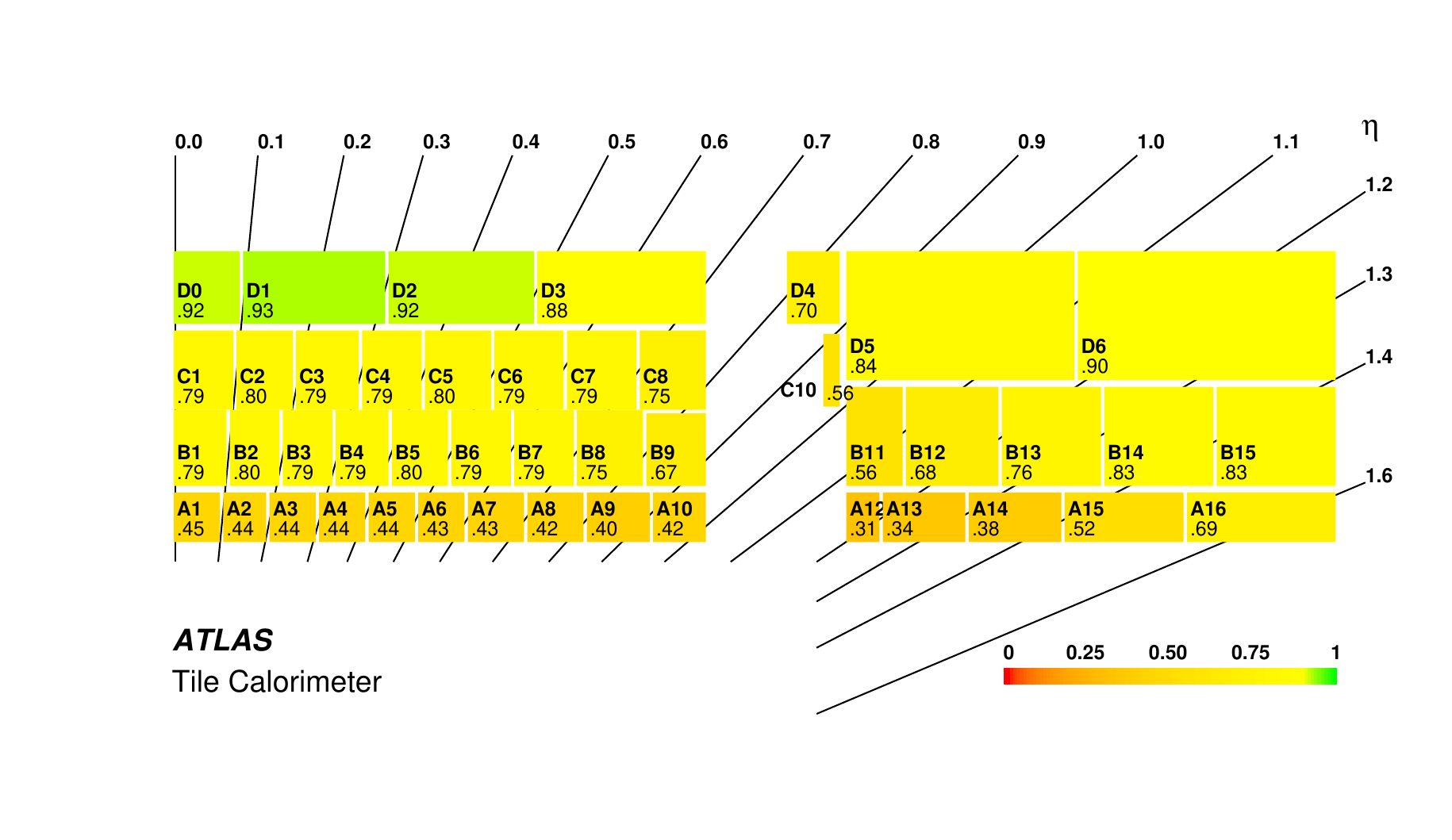}\hfill
\caption{Expected relative light yield $I/I_0$ of the TileCal barrel cells at the end of the HL-LHC, assuming an integrated luminosity of 4000~fb$^{-1}$ and a dose rate seven times higher than in Run~2 and Run~3. The relative uncertainties on the determinations are around 50\%.}
\label{fig:II0mapHL}
\end{figure}

Figure~\ref{fig:II0mapHL} presents the results of the $I/I_0$ extrapolation to the end of HL-LHC (4000~\ifb), including the effects of the anticipated higher dose rate. Results are also summarised in Table~\ref{tab:extrapolationR3HL}. The TileCal cells located in the B/BC- and D-layers are expected to have their light output reduced by no more than 25\%, except for B9, C10, B11 and B12 (loss up to around 40\%) which instrument the region close to the gap/crack and hence are more poorly shielded from radiation hazard. The A-layer is foreseen to be the most damaged, losing around half of the light response in the LB and attaining the extreme light reduction of 69\% for A12. The relative uncertainty on this extrapolation is around 50\%.

\subsection{Impact of Light signal degradation in the Calorimeter performance}
\label{sec:performance}

The light response of the TileCal is directly tied to the quality of its energy measurements. The loss of light signal is mainly compensated through calibration, stabilising the energy response scale, but impact the detector response uniformity, an effect which scales with the amount of degradation. On the other hand, the deterioration of the signal-to-noise ratio may negatively affect the calorimeter energy resolution. These factors contribute to a corrosion of the overall detector performance in the reconstruction of isolated particles and particle showers.

The TileCal energy response to isolated particles was studied in Run~2 analysing $pp$ collision events containing muons from $W\to\mu\nu$ decays and single hadrons~\cite{TCAL-2021-01}. For isolated muons, the energy deposit per unit length is evaluated and compared to simulation for the different cells and radial layers over time. The study allows to verify the detector calibration against response variations, namely those arising from optics damage. For single isolated hadrons, the ratio between the calorimeter energy measurement and the track momentum is compared to simulation. The results show the good uniformity of the detector response in $\phi$ and a stability of the calibrated detector response within 2\%, proving no impact of the Run~2 optics degradation.

The impact of the expected light loss in the future detector performance was studied by simulation~\cite{ATLAS-TDR-28}. An extreme reduction of 50\% in the cell light yield, to 35~p.e./GeV from the typical $\sim$70~p.e./GeV, was set while simulating the detection of muons impinging the detector at different pseudorapidity ranges. The loss of photo-statistics broadens the distribution of the muon energy reconstruction but not significantly. The consequence of this energy resolution worsening in the efficiency of detecting muons with energy above the cell pile-up noise threshold is also evaluated. For A-cells, the efficiency reduces by around 5\% (from 70\%), but is unaffected for D-cells due to small pile-up contribution. Overall, the simulation of an extreme scenario with 50\% light output shows no significant impact for muons.


%

%
%
\FloatBarrier

\section{Conclusion}
\label{sec:conclusion}

This paper presents a study of the radiation damage to the plastic optical material of the Tile Calorimeter during the LHC Run~2 operation.
Data from the calibration systems were explored to measure the light output lost by the calorimeter cells due to response degradation of scintillating tiles and WLS fibres as a function of time.
The maximum degradation in barrel cells was observed in the A-layer, which ended Run~2 with a 10\% loss at the most.
The degradation in the other layers is at or below the 1\% precision of the measurement.
The gap scintillators exhibited a total 12\% light response loss and the crack counters were replaced for the Run~3 after a total damage of 20\% and 30\% for E3 and E4 cells, respectively. MBTS inner and outer counters faced the most extreme doses and were also replaced for Run~3. Inner counters lost nearly 90\% of the light signal after 10~kGy of accumulated dose during Run~2, while outer counters lost almost totally their response after integrating 10 times more dose.
Moreover, the radiation damage to individual tiles was probed by extracting the signal amplitude for each tile from the $^{137}$Cs scans, with a precision of about 2\%. The study shows that the cell response non-uniformity in the EBA partition was kept between 6\% and 9\%, within the TileCal specifications. Prolonged radiation exposure is expected to increase non-uniformity in the future.
Finally, a simulation of the Run~2 radiation environment in the calorimeter and the expected accumulated doses in future runs was employed to model the radiation resistance of the TileCal barrel cells as a function of the dose and dose rate. Dose rate effects were found to have a significant impact on the degradation rate, in agreement with previous results from the CMS Hadron Endcap Calorimeter. Extrapolations based on the degradation model including the effect of a dose rate seven times higher in the HL-LHC indicate that the less exposed cells of the B/BC- and D-layers will degrade by no more than 25\% until the end of the HL-LHC phase (4000~\ifb). The A-layer is foreseen to be the most damaged, losing around half of the light response. The relative uncertainty of this prediction is around 50\%.


%

%
\section*{Acknowledgements}

%
%

%
%

The authors would like to acknowledge the entire TileCal community for their contribution with the discussions related to this work,
the operations acquiring calibration data, the input from the data quality activities, and the careful review of this report.
The authors extend their acknowledgement to the engineers and technical staff involved in the construction of the TileCal and to
the ATLAS radiation simulation group.


%
%
%
%
%

%
%
%
%
%
%

%
%
%

%
%
%
\printbibliography

\end{document}

%% file: atlas_authlist.tex
 
\begin{flushleft}
\hypersetup{urlcolor=black}

\bigskip

\AtlasOrcid{J.~Abdallah}$^\textrm{\scriptsize 1}$,
\AtlasOrcid[0000-0002-4355-5589]{M.N.~Agaras}$^\textrm{\scriptsize 6}$,
\AtlasOrcid[0000-0001-8638-0582]{A.~Ahmad}$^\textrm{\scriptsize 4}$,
\AtlasOrcid{G.~Arabidze}$^\textrm{\scriptsize 22}$,
\AtlasOrcid[0000-0003-1419-3213]{P.~Bartos}$^\textrm{\scriptsize 3}$,
\AtlasOrcid[0000-0002-1976-5703]{A.~Berrocal Guardia}$^\textrm{\scriptsize 2}$,
\AtlasOrcid[0000-0003-2138-9062]{D.~Bogavac}$^\textrm{\scriptsize 4}$,
\AtlasOrcid[0000-0003-1990-2947]{F.~Carrio Argos}$^\textrm{\scriptsize 18}$,
\AtlasOrcid[0000-0002-5567-4278]{L.~Cerda Alberich}$^\textrm{\scriptsize 18}$,
\AtlasOrcid[0000-0002-5376-2397]{B.~Chargeishvili}$^\textrm{\scriptsize 16}$,
\AtlasOrcid[0000-0002-9187-7478]{P.~Conde Mui\~no}$^\textrm{\scriptsize 13}$,
\AtlasOrcid[0000-0002-3279-3370]{A.~Cortes-Gonzalez}$^\textrm{\scriptsize 4}$,
\AtlasOrcid[0000-0002-5940-9893]{A.~Gomes}$^\textrm{\scriptsize 13}$,
\AtlasOrcid[0000-0002-3770-8307]{T.~Davidek}$^\textrm{\scriptsize 15}$,
\AtlasOrcid[0000-0002-9414-8350]{T.~Djobava}$^\textrm{\scriptsize 16}$,
\AtlasOrcid[0000-0003-4157-592X]{A.~Durglishvili}$^\textrm{\scriptsize 16}$,
\AtlasOrcid[0000-0002-4095-4808]{S.~Epari}$^\textrm{\scriptsize 2}$,
\AtlasOrcid[0000-0002-4056-4578]{G.~Facini}$^\textrm{\scriptsize 5}$,
\AtlasOrcid[0000-0003-4278-7182]{J.~Faltova}$^\textrm{\scriptsize 15}$,
\AtlasOrcid[0000-0002-5070-2735]{L.~Fiorini}$^\textrm{\scriptsize 18}$,
\AtlasOrcid{M.~Fontes Medeiros}$^\textrm{\scriptsize 4}$,
\AtlasOrcid{S. Fracchia}$^\textrm{\scriptsize 2}$,
\AtlasOrcid[0000-0003-3078-0733]{J.~Glatzer}$^\textrm{\scriptsize 5}$,
\AtlasOrcid[0000-0003-4315-2621]{A.J.~Gomez Delegido}$^\textrm{\scriptsize 18}$,
\AtlasOrcid[0000-0002-0309-4490]{S.~Harkusha}$^\textrm{\scriptsize 20}$,
\AtlasOrcid{A.M.~Henriques Correia}$^\textrm{\scriptsize 4}$,
\AtlasOrcid[0000-0002-8340-9455]{M.~Kholodenko}$^\textrm{\scriptsize 13}$,
\AtlasOrcid[0000-0003-1661-6873]{P.~Klimek}$^\textrm{\scriptsize 4}$,
\AtlasOrcid[0000-0002-9211-9775]{I.~Korolkov}$^\textrm{\scriptsize 2}$,
\AtlasOrcid[0000-0001-9099-0009]{A.~Maio}$^\textrm{\scriptsize 13}$,
\AtlasOrcid[0000-0003-2965-7746]{F.M.~Pedro Martins}$^\textrm{\scriptsize 13}$,
\AtlasOrcid[0000-0002-7006-0864]{J.G.~Saraiva}$^\textrm{\scriptsize 13}$,
\AtlasOrcid[0000-0002-8186-4032]{S.~Menke}$^\textrm{\scriptsize 9}$,
\AtlasOrcid[0000-0002-0654-8398]{K.~Mihule}$^\textrm{\scriptsize 15}$,
\AtlasOrcid[0000-0002-4688-3510]{I.A.~Minashvili}$^\textrm{\scriptsize 16}$,
\AtlasOrcid[0000-0003-2028-1930]{M.~Mlynarikova}$^\textrm{\scriptsize 4}$,
\AtlasOrcid[0000-0002-1720-0493]{M.~Mosidze}$^\textrm{\scriptsize 16}$,
\AtlasOrcid{N.~Mosulishvili}$^\textrm{\scriptsize 17}$,
\AtlasOrcid[0000-0001-8978-7150]{S.~Nemecek}$^\textrm{\scriptsize 14}$,
\AtlasOrcid[0000-0002-7139-9587]{R.~Pedro}$^\textrm{\scriptsize 13*}$,
\AtlasOrcid[0000-0002-9639-7887]{B.C.~Pinheiro Pereira}$^\textrm{\scriptsize 13}$,
\AtlasOrcid[0000-0001-5435-497X]{V.~Pleskot}$^\textrm{\scriptsize 15}$,
\AtlasOrcid[0000-0002-9929-9713]{S.~Polacek}$^\textrm{\scriptsize 15}$,
\AtlasOrcid[0000-0002-6960-502X]{Y.~Qin}$^\textrm{\scriptsize 2}$,
\AtlasOrcid{V.~Rossetti}$^\textrm{\scriptsize 21}$,
\AtlasOrcid[0000-0002-9095-7142]{R.~Rosten}$^\textrm{\scriptsize 2}$,
\AtlasOrcid[0000-0003-1710-9291]{H.~Santos}$^\textrm{\scriptsize 13}$,
\AtlasOrcid[0000-0002-8637-6134]{D.~Schaefer}$^\textrm{\scriptsize 5}$,
\AtlasOrcid[0000-0001-9569-3089]{F.~Scuri}$^\textrm{\scriptsize 12}$,
\AtlasOrcid{Y~Smirnov}$^\textrm{\scriptsize 10}$,
\AtlasOrcid[0000-0002-0518-4086]{C.A.~Solans Sanchez}$^\textrm{\scriptsize 4}$,
\AtlasOrcid[0000-0002-2737-8674]{A.A.~Solodkov}$^\textrm{\scriptsize 19}$,
\AtlasOrcid[0000-0002-2598-5657]{O.V.~Solovyanov}$^\textrm{\scriptsize 6}$,
\AtlasOrcid[0000-0002-9776-5880]{A.~Valero}$^\textrm{\scriptsize 18}$,
\AtlasOrcid[0000-0002-8483-9502]{H.G.~Wilkens}$^\textrm{\scriptsize 4}$,
\AtlasOrcid[0000-0001-7909-4772]{T.~Zakareishvili}$^\textrm{\scriptsize 16}$.
\bigskip
\\

$^{1}$Department of Physics, University of Texas at Arlington, Arlington TX; United States of America.\\
$^{2}$Institut de F\'isica d'Altes Energies (IFAE), Barcelona Institute of Science and Technology, Barcelona; Spain.\\
$^{3}$Faculty of Mathematics, Physics and Informatics, Comenius University, Bratislava; Slovak Republic.\\
$^{4}$CERN, Geneva; Switzerland.\\
$^{5}$Enrico Fermi Institute, University of Chicago, Chicago IL; United States of America.\\
$^{6}$LPC, Universit\'e Clermont Auvergne, CNRS/IN2P3, Clermont-Ferrand; France.\\
$^{7}$Department of Physics and Astronomy, University College London, London; United Kingdom.\\
$^{8}$Group of Particle Physics, University of Montreal, Montreal QC; Canada.\\
$^{9}$Max-Planck-Institut f\"ur Physik (Werner-Heisenberg-Institut), M\"unchen; Germany.\\
$^{10}$Department of Physics, Northern Illinois University, DeKalb IL; United States of America.\\
$^{11}$Ohio State University, Columbus OH; United States of America.\\
$^{12}$Dipartimento di Fisica E. Fermi, Universit\`a di Pisa, Pisa; Italy.\\
$^{13}$Laborat\'orio de Instrumenta\c{c}\~ao e F\'isica Experimental de Part\'iculas - LIP, Lisboa; Portugal.\\
$^{14}$Institute of Physics of the Czech Academy of Sciences, Prague; Czech Republic.\\
$^{15}$Charles University, Faculty of Mathematics and Physics, Prague; Czech Republic.\\
$^{16}$High Energy Physics Institute, Tbilisi State University, Tbilisi; Georgia.\\
$^{17}$University of Georgia, Tbilisi; Georgia.\\
$^{18}$Instituto de F\'isica Corpuscular (IFIC), Centro Mixto Universidad de Valencia - CSIC, Valencia; Spain.\\
$^{19}$School of Physics, University of the Witwatersrand, Johannesburg; South Africa.\\
$^{20}$Yerevan Physics Institute, Yerevan; Armenia.\\
$^{21}$Stockholm University, Stockholm, Sweden.\\
$^{22}$Michigan State University, East Lansing, MI, United States of America.\\
\bigskip
$^{*}$Corresponding author.
E-mail: \href{mailto:rute.pedro@cern.ch}{rute.pedro@cern.ch}

\end{flushleft}
